\documentclass[aps,preprint,floatfix,showpacs,superscriptaddress]{revtex4}

\usepackage{graphicx}
\usepackage{amssymb}
\usepackage{amsfonts}

\usepackage{bm}

\newcommand{\be}{\begin{equation}}
\newcommand{\ee}{\end{equation}}
\newcommand{\bel}[1]{\be\label{#1}}
\newcommand{\re}[1]{Eq.~(\ref{#1})}
\newcommand{\ds}{\displaystyle}
\newcommand{\ov}[1]{\overline{#1}}
\newcommand{\hsp}{\hspace*{1pt}}

\begin{document}

\title{
Equation of state of hadron resonance gas and\\
the phase diagram of strongly interacting matter}

\author{L.M. Satarov}

\affiliation{Frankfurt Institute for Advanced Studies,
J.W.~Goethe University, D--60438 Frankfurt
am Main, Germany}

\affiliation{The Kurchatov Institute, Russian Research Center,
123182 Moscow, Russia}

\author{M.N.~Dmitriev}

\affiliation{The Kurchatov Institute, Russian Research Center,
123182 Moscow, Russia}

\author{I.N.~Mishustin}

\affiliation{Frankfurt Institute for Advanced Studies,
J.W.~Goethe University, D--60438 Frankfurt
am Main, Germany}

\affiliation{The Kurchatov Institute, Russian Research Center,
123182 Moscow, Russia}

\begin{abstract}
The equation of state of hadron resonance gas at finite temperature and baryon density
is calculated taking into account finite-size effects within the excluded volume
model. Contributions of known hadrons with masses up to 2 GeV are included
in the zero-width approximation.
Special attention is paid to the role of strange hadrons in the system with zero
total strangeness. A density--dependent mean field is added to guarantee that
the nuclear matter has a saturation point and a~liquid-gas phase transition.
The deconfined phase is described by the bag model with lowest order perturbative corrections.
The phase transition boundaries are found by using the Gibbs conditions with the strangeness
neutrality constraint. The sensitivity of the phase diagram to the hadronic
excluded volume and to the parametrization of the mean--field is investigated.
The possibility of strangeness--antistrangeness separation in the mixed phase is analyzed.
It is demonstrated that the peaks in the $K/\pi$ and $\Lambda/\pi$ excitation functions
observed at low SPS energies can be explained by a nonmonotonous behavior of the strangeness
fugacity along the chemical freeze-out line.
\end{abstract}


\maketitle

\section{Introduction}

Relativistic heavy--ion collisions represent a powerful tool for studying properties
of strongly interacting matter in the laboratory. The main goal is to explore
the phase diagram of such matter and, in particular, to
investigate properties of a new phase -- the quark-gluon plasma (QGP).
This can be done only indirectly, i.e. by comparing predictions of
different theoretical models with experimental data. The fluid-dynamical
model is one of the most popular models for describing the relativistic heavy--ion collisions.
In this model one needs the equation of state (EOS) and transport coefficients
as input information. The QCD lattice calculations can give reliable results on
thermodynamic properties of strongly interacting matter only for small baryon
chemical potentials. In this case a crossover type of the deconfinement
transition is predicted~\cite{Kar01}. It seems that this prediction is confirmed
by recent RHIC experiments (see e.g.~\cite{Gyu04}). The structure
of the phase diagram at high baryon densities remains rather uncertain~\cite{Phi07}.
However, exactly the baryon-rich matter attracts the
main interest due to the possibility of the first order deconfinement phase
transition~~\cite{Alf98,Sca01}. The future facility for antiproton and ion research
(FAIR) at GSI (Darmstadt) is especially focused on heavy-ion collisions with bombarding
energies 20-40 AGeV, where maximal baryon densities of about 10~times the normal
nuclear density $n_0=0.15$ fm$^{-3}$ are expected~\cite{Ars07}. Some interesting
results concerning dense baryon matter have been
obtained already at low SPS energies by the NA49 Collaboration~\cite{Afa02,Alt08}.

To understand properties of the deconfinement transition, one needs to
know accurately the EOS of the hadronic as well as the quark--gluon phases.
As demonstrated in Ref.~\cite{Mis02}, very often arbitrarily chosen models for these two phases
do not lead to any phase transition. For instance, taking an ideal resonance gas for
the hadronic phase and the bag model for the~QGP, one comes to the paradoxical
conclusion that the hadronic phase is thermodynamically stable at high
temperatures~\cite{Hei86}.
It is clear that at high densities the repulsive interaction between hadrons becomes
important and should be explicitly taken into account.
One can introduce such interaction via the vector meson exchange as done in
the relativistic mean--field models of the Walecka type~\cite{Wal74}. However,
the repulsive vector field in these models is proportional to the net baryon
density and, therefore, can not solve the problem in the case of baryon--free matter.
Another possibility is to take into account
the finite size of hadrons within the Van der Waals approach~\cite{Lan80}. Different
versions of the
EOS of strongly interacting matter with finite--size corrections have been considered
in \mbox{Refs.~\cite{Hag80,Kar80,Kap81,Ris91,Ven92,Yen97,Whe04}}, but the sensitivity
of the phase diagram to the choice of hadronic volumes was not fully
investigated. In this paper we address this problem on the quantitative level.
This allows us to derive a consistent EOS with the deconfinement phase transition.
We also study the role of the
strangeness neutrality constraint, which previously has been
investigated only qualitatively~\cite{Gre87,Gor05}.

The paper is organized as follows. In Sect.~II we describe thermodynamic
properties of a hadron resonance gas in the excluded volume approximation.
In this section we show that the strangeness
neutrality constraint leads to a non-monotonic behavior of the $K/\pi$ and $\Lambda/\pi$
multiplicity ratios along the freeze--out line in the chemical potential--temperature plane.
The bag model for a quark--gluon phase is formulated in Sect.~III.
The phase diagram of strongly interacting matter is studied in Sect.~IV.
In particular, we investigate its sensitivity to the choice of hadronic volumes.
In Sect.~V we introduce the mean--field interaction
of baryons in order to implement the liquid--gas phase transition in nuclear matter
at low temperatures. In Sect.~VI we summarize our results and outline possible
improvements of the model.

\section{Hadronic phase\label{FORM}}

\subsection{Hadron resonance gas within the excluded volume approximation\label{HEOS}}

Let us consider a purely hadronic system in the total
volume $V$ assuming local thermodynamic equilibrium and neglecting the isospin,
Coulomb and surface effects. Since hadrons are composite particles, their finite sizes
should be taken into account at high enough densities or temperatures.
The intuitive way to implement the finite-size effects is to use the
Van der Waals prescription which is known also as the excluded volume approximation.
Within such an approach the hadronic system is still regarded
as an ideal gas, but in the volume reduced by the volume occupied by constituents,
\bel{vrep}
V^{\hsp\prime}  =V-\sum_i v_i N_i\hsp.
\ee
Here $v_i$ and $N_i$ are the excluded volume and the number of hadrons of type $i$\hsp,
the sum runs over all hadronic species $i$\hsp. In the following we assume
that all hadrons have the same radius~$r_h$. Then one can estimate the excluded volume
per particle as 1/2 of a spherical volume with the radius $2r_h$~\cite{Lan80}
\bel{exv}
v_i=v=\frac{16\pi}{3}\hsp r_h^3\,.
\ee
This approximation should be good enough at particle densities much smaller than
the density of close packing\hsp\footnote
{
As claimed in Ref.~\cite{Kar80}, crystallization into a solid state should occur
at densities near the close packing limit
\mbox{$n_{\rm cp}=(4\sqrt{2}\,r_h^3)^{-1}\simeq 3\hsp v^{-1}$}.
Our estimates show, however, that a dense hadronic
system should earlier undergo the transition into a quark--gluon phase.
}.
The quantity $v$ will be considered below as a model parameter.
According to the analysis of hadronic yields observed in relativistic heavy--ion
collisions~\cite{Yen97}, the reasonable interval of hadronic radii is $r_h\simeq (0.3-0.6)$ fm.
This corresponds to the excluded volume range $v\simeq (0.5-3)$ fm$^3$.

Making the replacement $V\to V^{\hspace*{0.5pt}\prime}$ in the canonical partition sum,
one obtains the following equation for the pressure of hadronic system~\cite{Ris91}
\bel{pres}
P=\sum\limits_i P_i\hsp(\widetilde{\mu}_i, T)\,,
\ee
where $P_i\hsp(\widetilde{\mu_i},T)$ is the partial pressure of ideal gas of $i$-th hadrons \
at temperature $T$ and chemical potential $\widetilde{\mu_i}$\,. The relation connecting
$\widetilde{\mu_i}$ with the real chemical potential $\mu_i$ reads
\bel{cpti}
\widetilde{\mu_i}=\mu_i-v\hsp P.
\ee
In fact, Eqs.~(\ref{pres})--(\ref{cpti}) give the integral equation for $P$ at a given
temperature and set of chemical potentials $\{\mu_i\}$\,. In the limiting case \mbox{$v\to 0$} one
obtains the Dalton law for a~mix\-ture of ideal gases of different hadrons.

The condition of chemical equilibrium with respect to strong interactions leads to the
relations
\bel{cceq}
\mu_i=B_i\hsp\mu+S_i\hsp\mu_S\,,
\ee
where $B_i=0\hsp,\pm\hsp 1$ and $S_i=0,\pm\hsp 1\hsp, \pm\hsp 2\hsp\ldots$ are,
respectively, the baryon and
strangeness quantum numbers of the $i$--th hadronic species, $\mu$ and $\mu_S$ denote
the baryon and
strange chemical potentials. Therefore, pressure of a chemically equilibrated system
is a function of only three independent variables $\mu,\mu_S,T$:
\bel{pres1}
P=P\hsp (\mu,\mu_S,T)\hsp .
\ee
This function satisfies the thermodynamic relation~\cite{Lan80}
\bel{ther}
dP=s\hsp dT+n\hspace*{0.5pt} d\mu+n_S\hsp d\mu_S\,,
\ee
where $s$ is the entropy density, $n$ and $n_S$ are, respectively, the baryon and
strangeness number densities. In this paper we consider only the hadronic matter with zero net
strangeness. Such matter may be created e.g. in collisions of nonstrange projectile and target
nuclei. Since strangeness is conserved in strong interactions, the produced system must obey
the condition of strangeness neutrality
\bel{strd}
n_S\hsp (\mu,\mu_S,T)=0\,.
\ee
In a baryon--asymmetric matter ($\mu\ne 0$) this condition can be satisfied only at nonzero
strange chemical potential $\mu_S$\hsp. It easy to show that $\mu_S$ should be positive
in the baryon-rich matter. This follows from the fact that the negative strangeness
is carried by hyperons while the positive strangeness is associated mostly with kaons.

From Eqs. (\ref{pres})--(\ref{cceq}), and (\ref{ther}), one can obtain
explicit relations for thermodynamic quantities $s,n,n_S$~\cite{Yen97}:
\begin{eqnarray}
&&s=\frac{\partial P}{\partial T}=r\sum\limits_i
\widetilde{s}_i\hsp(\widetilde{\mu}_i, T)\,,\label{entd}\\
&&n=\frac{\partial P}{\partial\mu}=r\sum\limits_i
B_i\hsp\widetilde{n}_i\hsp(\widetilde{\mu}_i, T)\,,\label{bden}\\
&&n_S=\frac{\partial P}{\partial\mu_S}=r\sum\limits_i
S_i\widetilde{n}_i\hspace*{1.5pt}(\widetilde{\mu}_i, T)\,,\label{strd1}
\end{eqnarray}
where $\widetilde{s}_i=\partial P_i/\partial T$ and
$\widetilde{n}_i=\partial P_i/\partial\widetilde{\mu}_i$
are, respectively, the entropy and particle number density of the ideal gas of $i$-th
hadrons. The reduction factor $r$ in r.h.s of Eqs. (\ref{entd})--(\ref{strd1})
is defined as
\bel{rfac}
r=(1+v\sum\limits_i\widetilde{n}_i)^{-1}.
\ee
The number density of the $i$--th species satisfies the relations
\bel{pard}
n_i=\frac{\partial P}{\partial\mu_i}=r\hsp\widetilde{n}_i\leqslant\frac{1}{v}\,.
\ee

By using Eqs. (\ref{pres})--(\ref{cceq}), (\ref{entd})--(\ref{rfac}) and the Gibbs--Duhem equation
\bel{gdeq}
\epsilon=-P+Ts+\mu\hsp n+\mu_S n_S\,,
\ee
one obtains the expression for energy density
\bel{ende}
\epsilon=r\sum\limits_i\widetilde{\epsilon}_i\hsp(\widetilde{\mu}_i, T)\,,
\ee
where $\widetilde{\epsilon}_i=-P_i+T\hsp\widetilde{s}_i+\widetilde{\mu}_i\hsp\widetilde{n}_i$
is energy density of the ideal gas of $i$-th hadrons.
Below we take into account the contributions of stable hadrons as well as
mesonic and baryonic resonances in the zero-width approximation.
In  the grand canonical ensemble the thermodynamic functions of the relativistic
ideal gas of species $i$ are given by the following explicit rela\-tions~($\hbar=c=1$)
\bel{therf}
\left(\begin{array}{c}
\widetilde{\epsilon}_i\\P_i\\
\widetilde{n}_i
\end{array}\right)=\frac{g_i}{2\hsp\pi^2}\int\limits_{m_i}^{\infty}d\epsilon\hsp
\frac{\sqrt{\epsilon^2-m_i^2}}{\exp{\left(\frac{\ds\epsilon-\widetilde{\mu}_i}
{\raisebox{-3pt}{$T$}}\right)}\pm 1}
\left(\begin{array}{c}
\epsilon^2\\
\frac{\ds 1}{\raisebox{-3pt}{$\ds 3$}}\hsp (\epsilon^2-m_i^2)\\
\epsilon
\end{array}\right)\hsp,
\ee
where $m_i$ is the mass of the $i$-th hadron and $g_i$ is its spin--isospin degeneracy
factor. The lower sign in r.h.s. of (\ref{therf}) corresponds to mesons ($B_i=0$)\label{fref1}
\hsp\footnote
{
In Eq.~(\ref{therf}) we assume that the condition of Bose--condensation
$\widetilde{\mu}_i>m_i$ is not satisfied (see below).
}
and the upper one, to baryons ($B_i=1$) or antibaryons ($B_i=-1$). The
integrals in Eq.~(\ref{therf}) are calculated numerically as explained in Appendix.

\renewcommand{\baselinestretch}{0.5}
\begin{table}
\caption{Characteristics of hadronic species included in the
calculation\hsp\protect\footnote{
Values with asterisk correspond to antihyperons.
}.
}
\label{tab1}
\footnotesize
\begin{ruledtabular}
\begin{tabular}{ccrcrcll|ccrcrcll}
hadron & $m_i$ (GeV) & $g_i$ & $B_i$ & $S_i$ & $I_i$ & $d_i^\pi$
&$d_i^K$ & hadron & $m_i$ (GeV) & $g_i$ & $B_i$ & $S_i$ &
$I_i$&$d_i^\pi$ & $d_i^K$ \\[1pt]
\hline
$\pi$             & 0.140 & 3  & 0 & 0  & 1   &  1 &  0 & $N$\hsp(1535)     & 1.530 & 4  & 1 & 0  & 1/2~~&1.45~&0\\
$K$               & 0.496 & 2  & 0 & 1  &1/2~~&  0 &  1 & $\pi_1$\hsp(1600) & 1.596 & 9  & 0 & 0  & 1  &3.54&0.03\\
$\ov{K}$          & 0.496 & 2  & 0 &$-1$& 1/2 &  0 &  0 & $\Delta$\hsp(1600)& 1.600 & 16 & 1 & 0  & 3/2&1.93&  0 \\
$\eta$            & 0.543 & 1  & 0 & 0  & 0   &1.95~& 0 & $\Lambda$\hsp(1600)&1.600 & 2  & 1 &$-1$& 0  &0.6&0.4* \\
$\rho$            & 0.776 & 9  & 0 & 0  & 1   &  2 &  0 & $\Delta$\hsp(1620)& 1.630 &  8 & 1 & 0  & 3/2&1.75&  0 \\
$\omega$          & 0.782 & 3  & 0 & 0  & 0   &2.79&  0 & $\eta_2$\hspace*{1.5pt}(1645)&1.617&5&0&0& 0 &3.24& 0.1\\
$K^*$             & 0.892 & 6  & 0 & 1  & 1/2 &  1 &  1 & $N$\hsp(1650)     & 1.655 &  4 & 1 & 0  & 1/2&1.2 &0.07\\
$\ov{K}^{\hsp *}$ & 0.892 & 6  & 0 &$-1$& 1/2 &  1 &  0 & $\omega$\hsp(1650)& 1.670 &  3 & 0 & 0  & 0  &4.18&  0 \\
$N$               & 0.939 & 4  & 1 & 0  & 1/2 &  0 &  0 & $\Sigma$\hsp(1660)& 1.660 &  6 & 1 &$-1$& 1  &0.8 &0.2*\\
$\eta^{\hsp\prime}$     & 0.958 & 1  & 0 & 0  & 0   &3.25&  0 & $\Lambda$\hsp(1670)&1.670 & 2  & 1 &$-1$& 0  &0.89&0.3*\\
$f_0$             & 0.980 & 1  & 0 & 0  & 0   &2.95&  0 & $\Sigma$\hsp(1670)& 1.670 & 2  & 1 &$-1$& 1  &1.25&  0 \\
$a_0$             & 0.980 & 3  & 0 & 0  & 1   &2.95&0.15& $\omega_3$\hsp(1670)&1.667&  7 & 0 & 0  & 0  &3.90&  0 \\
$\phi$            & 1.020 & 3  & 0 & 0  & 0   &0.48&0.83& $\pi_2$\hsp(1670) & 1.672 & 15 & 0 & 0  & 1  &3.90&0.04\\
$\Lambda$         & 1.116 & 2  & 1 &$-1$& 0   &  0 &  0 & $\Omega^-$        & 1.672 &  4 & 1 &$-3$& 0  &0.32&0.68*\\
$h_1$             & 1.170 & 3  & 0 & 0  & 1   &  3 &  0 & $N$\hsp(1675)     & 1.675 & 12 & 1 & 0  & 1/2~~&1.6 & 0\\
$\Sigma$          & 1.189 & 6  & 1 &$-1$& 1   &  0 &  0 & $\phi$\hsp(1680)  & 1.680 & 3  & 0 & 0  & 0  &  1 & 0.5\\
$a_1$             & 1.230 & 9  & 0 & 0  & 1   &  3 &  0 & $K^*$\hsp(1680)   & 1.717 & 6  & 0 & 1  & 1/2&1.61&  1 \\
$b_1$             & 1.230 & 9  & 0 & 0  & 1   &3.79&  0 & $\ov{K}^{\hsp *}$\hsp(1680)&1.717&6&0&$-1$&1/2&1.61& 0 \\
$\Delta$          & 1.232 & 16 & 1 & 0  & 3/2 &  1 &  0 & $N$\hsp(1680)     & 1.685 & 12 & 1 & 0  & 1/2~~&1.35& 0\\
$f_2$             & 1.270 & 5  & 0 & 0  &  0  &2.20&0.05& $\rho_3$\hsp(1690)& 1.688 & 21 & 0 & 0  & 1  &3.35&0.05\\
$K_1$             & 1.273 & 6  & 0 & 1  & 1/2 &2.12&  1 & $\Lambda$\hsp(1690)&1.690 & 4  & 1 &$-1$& 0  &1.2 &0.25*\\
$\ov{K}_1$        & 1.273 & 6  & 0 &$-1$& 1/2 &2.12&  0 & $\Xi$\hsp(1690)  & 1.690 & 8  & 1 &$-2$& 1/2&0.33&0.66*\\
$f_1$             & 1.285 & 3  & 0 & 0  & 1   &3.69&0.09& $\rho$\hsp(1700)  & 1.720 & 9  & 0 & 0  & 1  &  4 &  0 \\
$\eta$\hsp (1295) & 1.295 & 1  & 0 & 0  & 0   &3.95&  0 & $N$\hsp(1700)     & 1.700 & 8  & 1 & 0  & 1/2~~&1.9& 0 \\
$\pi$\hsp(1300)   & 1.300 & 3  & 0 & 0  & 1   &  3 &  0 & $\Delta(1700)$     & 1.700 & 16 & 1 & 0  & 3/2&1.85&  0 \\
$\Xi$             & 1.315 & 4  & 1 &$-2$& 1/2 &  0 &  0 & $N$\hsp(1710)     & 1.710 & 4  & 1 & 0  & 1/2~~&1.65&0.15\\
$a_2$             & 1.318 & 15 & 0 &  0 & 1   &3.04&0.05& $f_0$\hsp(1710)   & 1.714 & 1  & 0 & 0  & 0  &2.87&0.25\\
$f_0$\hsp (1370)  & 1.370 & 1  & 0 & 0  & 1   &  2 &  0 & $N$\hsp(1720)     & 1.720 & 8  & 1 & 0  & 1/2~~&1.71&0.07\\
$\Sigma$\hsp(1385)& 1.385 & 12 & 1 &$-1$& 1   &  1 &  0 & $\Sigma$\hsp(1750)& 1.750 & 6  & 1 &$-1$& 1  &1.12&0.4*\\
$K_1$\hsp(1400)   & 1.400 & 6  & 0 & 1  & 1/2 &1.97&  1 & $K_2$\hsp(1770)   & 1.773 & 10 & 0 & 1  & 1/2&2.65&  1 \\
$\ov{K}_1$\hsp(1400)& 1.400& 6 & 0 &$-1$& 1/2 &1.97&  0 & $\ov{K}_2$\hsp(1770)&1.773& 10 & 0 &$-1$& 1/2&2.65&  0 \\
$\eta$\hspace*{1.5pt}(1405)&1.405&1& 0&0& 0   &2.59& 0.3& $\Sigma$\hsp(1775)& 1.775 & 18 & 1 &$-1$& 1  &1.58&0.45*\\
$\Lambda$\hsp(1405)& 1.406 & 2 & 1 &$-1$& 0   &  1 & 0  & $K_3^*$\hsp(1780) & 1.776 & 14 & 0 & 1  & 1/2&1.79&  1 \\
$K^*\hsp(1410)$    & 1.414 &  6 & 0 &  1 & 1/2 &1.54&  1 & $\ov{K}_3^{\hsp *}$\hsp(1780)&1.776&14&0&$-1$&1/2&1.79&0\\
$\ov{K}^{\hsp*}\hsp(1410)$& 1.414 &  6 & 0 &$-1$& 1/2 &1.54&  0 & $\pi$\hsp(1800)   & 1.812 & 3  & 0 & 0  & 1  &3.76&0.25\\
$f_1$\hsp(1420)   & 1.420 & 3  & 0 & 0  & 1   &  1 &  0 & $\Lambda$\hsp(1800)&1.800 & 2  & 1 &$-1$& 0  &0.92&0.54*\\
$\omega$\hsp(1420)& 1.420 & 3  & 0 & 0  & 0   &  3 &  0 & $\Lambda$\hsp(1810)&1.810 & 2  & 1 &$-1$& 0  &0.65&0.75*\\
$K_0^*$           & 1.425 &  2 & 0 &  1 & 1/2 &0.93&  1 & $K_2$\hsp(1820)   & 1.816 & 10 & 0 & 1  & 1/2&2.41&  1 \\
$\ov{K}_0^*$      & 1.425 &  2 & 0 &$-1$& 1/2 &0.93&  0 & $\ov{K}_2$\hsp(1820)&1.816& 10 & 0 &$-1$& 1/2&2.41&  0 \\
$K_2^*$           & 1.430 & 10 & 0 & 1  & 1/2 &1.65&  1 & $\Lambda$\hsp(1820)&1.820 & 6  & 1 &$-1$& 0  &0.69&0.89*\\
$\ov{K}_2^*$      & 1.430 & 10 & 0 &$-1$& 1/2 &1.65&  0 & $\Xi$\hsp(1820)  & 1.823 & 8  & 1 &$-2$& 1/2& 0  & 1* \\
$N$\hsp(1440)     & 1.440 & 4  & 1 & 0  & 1/2 &1.4 &  0 & $\Lambda$\hsp(1830)&1.830 & 6  & 1 &$-1$& 0  &1.31&0.45*\\
$\rho$\hsp(1450)  & 1.465 & 9  & 0 & 0  & 1   &  2 &  0 & $\phi_3$\hsp(1850)& 1.854 & 7  & 0 & 0  & 0  &0.5 &0.75\\
$a_0$\hsp(1450)   & 1.472 & 3  & 0 & 0  & 1   &2.38&0.33& $\pi_2$\hsp(1880) & 1.895 & 15 & 0 & 0  &  1 &4.85&  0 \\
$\eta$\hspace*{1.5pt}(1475)&1.476&1&0&0 & 0   &  1 &  1 & $\Lambda$\hsp(1890)&1.890 & 4  & 1 &$-1$& 0  &0.87&0.62*\\
$f_0$\hsp(1500)   & 1.505 & 1  & 0 & 0  & 0   &2.98&0.09& $\Delta(1905)$     & 1.890 & 24 & 1 & 0  & 3/2& 1.9&  0 \\
$\Lambda$\hsp(1520)&1.520 & 4  & 1 &$-1$& 0   &0.62&0.45* & $\Delta(1910)$     & 1.910 &  8 & 1 & 0  & 3/2&1.78&  0 \\
$N$\hsp(1520)     & 1.520 & 8  & 1 & 0  & 1/2 &1.4 &  0 & $\Delta(1920)$     & 1.920 & 16 & 1 & 0  & 3/2&1.88&  0 \\
$f'_2$\hsp(1525)  & 1.525 & 5  & 0 & 0  & 0   &0.42&0.89& $\Delta(1930)$     & 1.930 & 24 & 1 & 0  & 3/2& 1.9&  0 \\
$\Xi$\hsp(1530)   & 1.533 & 8  & 1 &$-2$& 1/2 &  1 &  0 & $\Delta(1950)$     & 1.930 & 32 & 1 & 0  & 3/2& 1.8&  0 \\
\end{tabular}
\end{ruledtabular}
\normalsize
\end{table}
\renewcommand{\baselinestretch}{1.2}

In the present work we include contributions of the lightest hadrons with masses
\mbox{$m_i\lesssim 2$\,GeV}. Altogether we take into account 59 mesonic and 41 baryonic species
listed in Ref.~\cite{PDG08}. This corresponds to 307 different isospin states of mesons, baryons and
antibaryons. Characteristics of hadrons included in our calculations are given in Table~\ref{tab1}.
For hadrons with nonzero isospin we use the isospin averaged mass values.
Note, that unless otherwise stated, we do not include a very broad scalar meson
resonance $f_0 (600)$ with mass
$m\sim 0.6$~GeV and width \mbox{$\Gamma\gtrsim 0.6$ GeV}\hsp\footnote
{
Except of $f_0 (600)$, a very similar set of hadrons has been used in the
THERMUS thermal model~\cite{Whe04}.
}.
The last two columns of
Table~\ref{tab1} give the average numbers of pions ($d_i^\pi$) and kaons ($d_i^K$) produced
in decays of the $i$--th hadron. These numbers are calculated by using the decay branching
ratios from Ref.~\cite{PDG08}\hsp\footnote{
For the observed decay channels with unknown probabilities we assume equal
branching ratios.
}.
For particles, stable with respect to strong interactions, we use the values $d_i^\pi=d_i^K=0$.
In decays of heavy hyperons only $\ov{K}$--mesons, not kaons, are produced.
In these cases the last column of Table~\ref{tab1} gives the contributions of corresponding
antihyperons (see entries with asterisks). For example, in the case $i=\Omega^-$ one has
$d_\Omega^K=0,\,d_{\ov{\Omega}}^K\simeq 0.68$.

\subsection{Thermodynamic properties of hadronic phase}

\begin{figure*}[htb!]
\centerline{\includegraphics[width=0.7\textwidth]{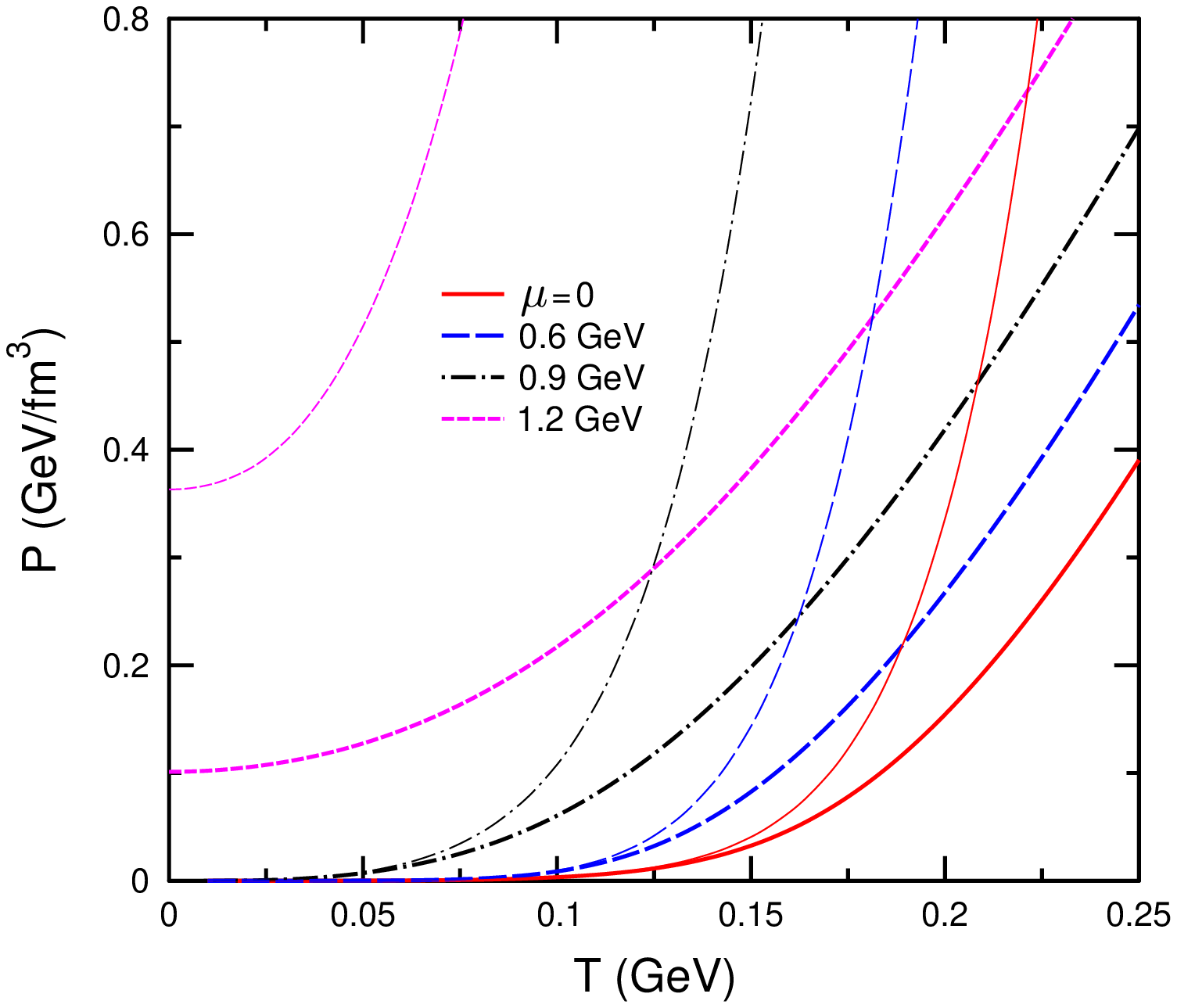}}
\caption{
Pressure of nonstrange hadronic system as a function of temperature
at different values of the baryon chemical potential $\mu$\hsp.
Thick and thin lines correspond to the excluded volume values $v=1$\,fm$^{3}$
and $v=0$, respectively.
}
\label{fig:t-p1}
\end{figure*}
\begin{figure*}[hbt!]
\centerline{\includegraphics[width=0.7\textwidth]{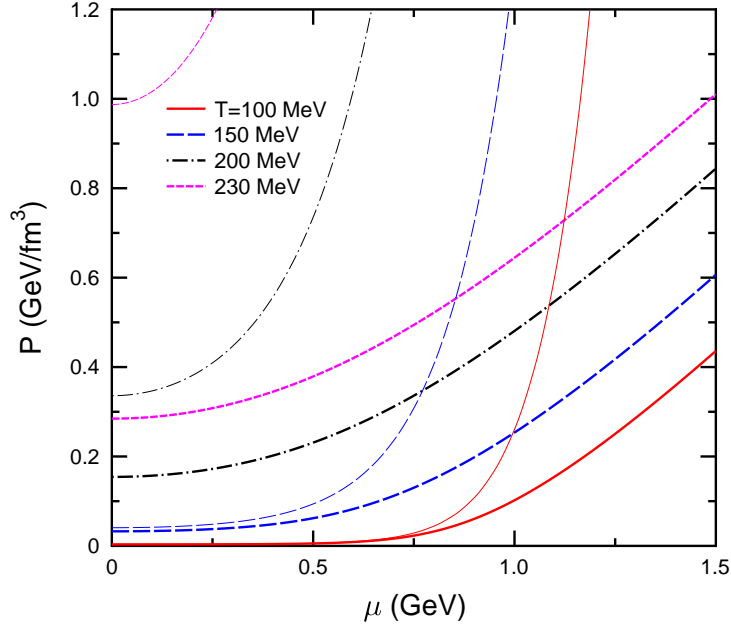}}
\caption{
Pressure isotherms for nonstrange hadronic system
as functions of the baryon chemical potential $\mu$\,.
Thick and thin lines correspond to values $v=1$\,fm$^{3}$
and $v=0$, respectively.
}
\label{fig:m-p2}
\end{figure*}
\begin{figure*}[hbt!]
\centerline{\includegraphics[width=0.7\textwidth]{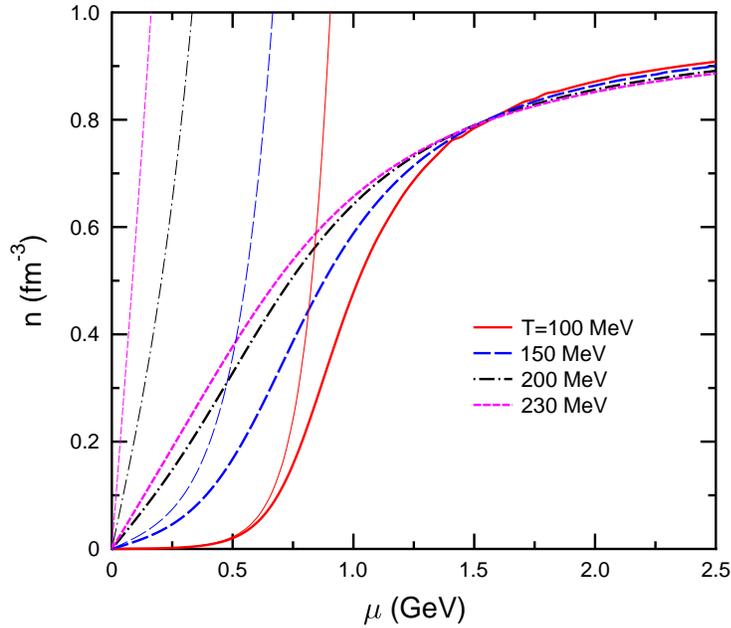}}
\caption{
Same as Fig.~\ref{fig:m-p2} but for isotherms of baryon density.
}
\label{fig:m-n3}
\end{figure*}

Below we present numerical results for a nonstrange ($n_S=0$), isospin-symmetric
hadronic matter. Figure~\ref{fig:t-p1} shows pressure $P$
as a function of $T$ for different values of $\mu$\hsp.
One can see that the excluded volume corrections (EVC) reduce the pressure as compared to
the calculation with $v=0$\hsp. The difference between two calculations increases with raising $T$
and~$\mu$\hsp. At~$\mu\gtrsim 1$~GeV  it is large for any $T$. Pressure isotherms
as functions of $\mu$ are shown in Fig.~\ref{fig:m-p2}\hsp.
In accordance with ~\re{bden}, smaller derivatives of the isotherms imply lower baryon densities
at $v\neq 0$ as compared with the case $v=0$\,. This is indeed seen in Fig.~\ref{fig:m-n3}.
As~expec\-ted, the saturating behavior $n\to 1/v$ is predicted at $\mu\to\infty$\hsp.

\begin{figure*}[hbt!]
\centerline{\includegraphics[width=0.7\textwidth]{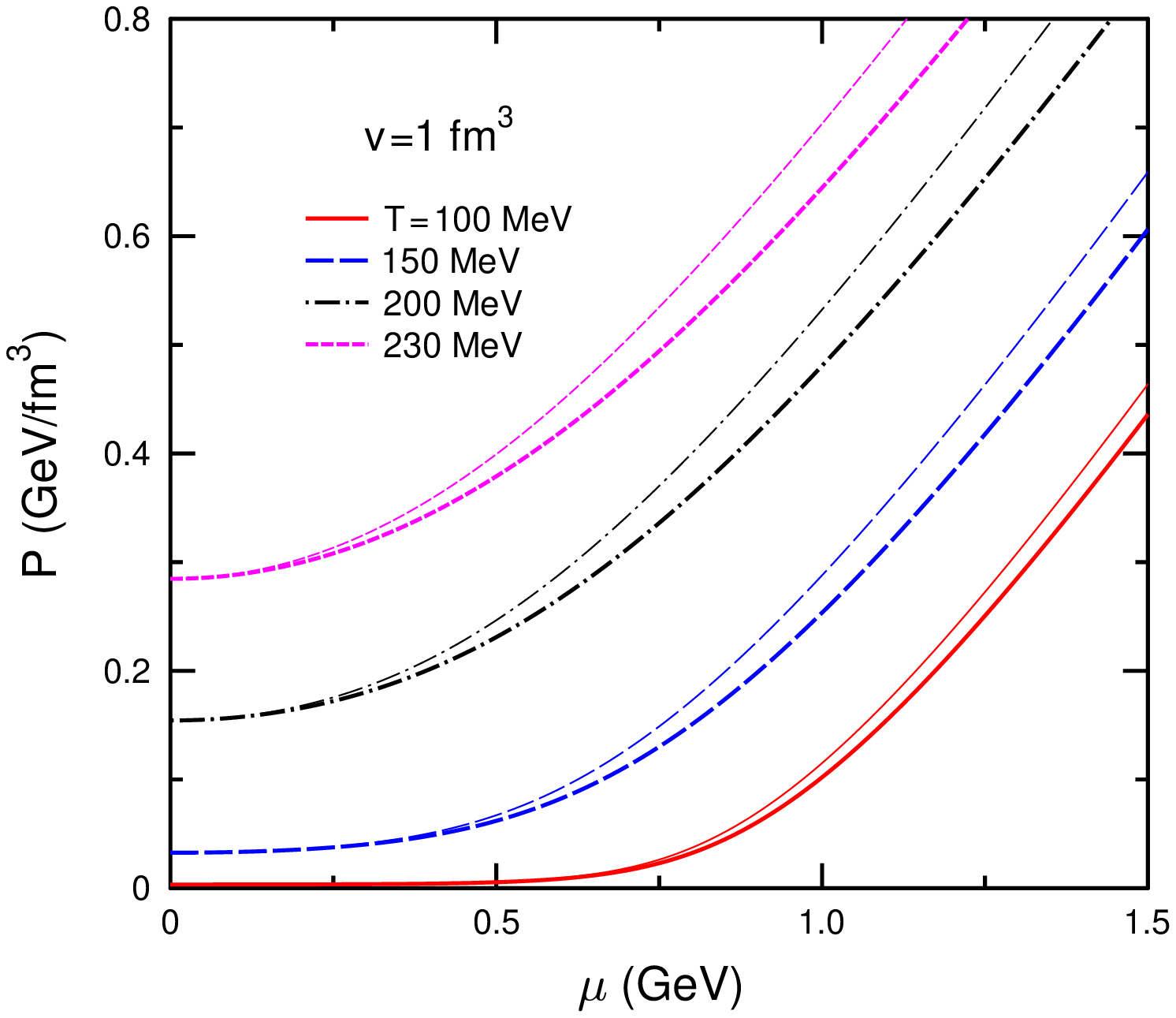}}
\caption{
Pressure isotherms of hadronic system. Thick and thin lines are calculated
assuming $n_S=0$ and $\mu_S=0$\hsp, respectively. All curves correspond to $v=1$~fm$^3$.
}
\label{fig:m-p4}
\end{figure*}

\begin{figure*}[hbt!]
\centerline{\includegraphics[width=0.7\textwidth]{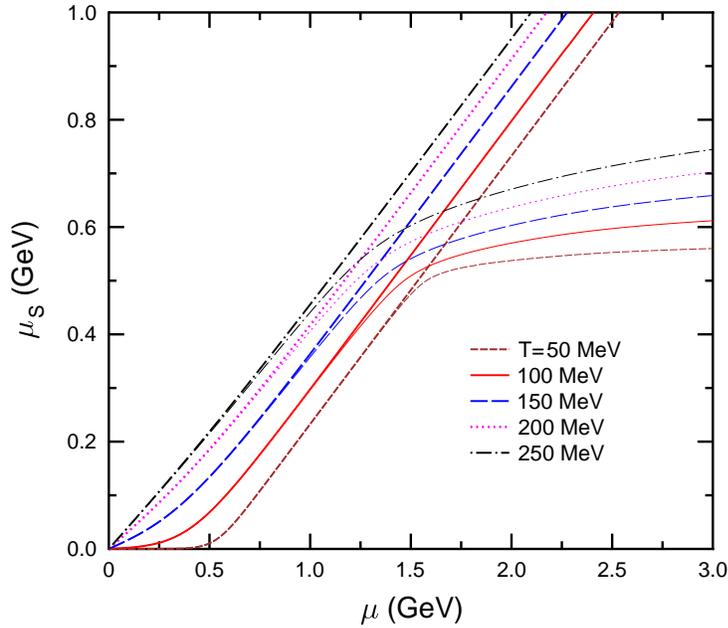}}
\caption{
Isotherms of $\mu_S$ for nonstrange hadronic system: thin and thick lines correspond to
$v=0$ and $v=1$ fm$^3$, respectively.
}
\label{fig:m-ms5}
\end{figure*}

To demonstrate sensitivity of the hadronic EOS to the strange chemical potential
$\mu_S$\hsp, in Fig.~\ref{fig:m-p4} we compare isotherms of pressure calculated assuming
either $n_S=0$ or $\mu_S=0$  (in the latter case the strangeness neutrality
is not guaranteed)\hsp. One can see that deviations caused by nonzero $\mu_S$
increase with $\mu$\hsp, but they are rather small, especially at low~$T$.
Comparison of $\mu_S$--isotherms as functions of $\mu$, calculated with and without EVC,
is shown in Fig.~\ref{fig:m-ms5}. For $v\neq 0$ the model predicts almost linear relation
between $\mu_S$ and $\mu$ with the slope of about~1/2.~\label{musref}
The deviations from the ideal gas of point-like hadrons ($v=0$) become important at $\mu\gtrsim 1$\,GeV.
\begin{figure*}[hbt!]
\centerline{\includegraphics[width=0.7\textwidth]{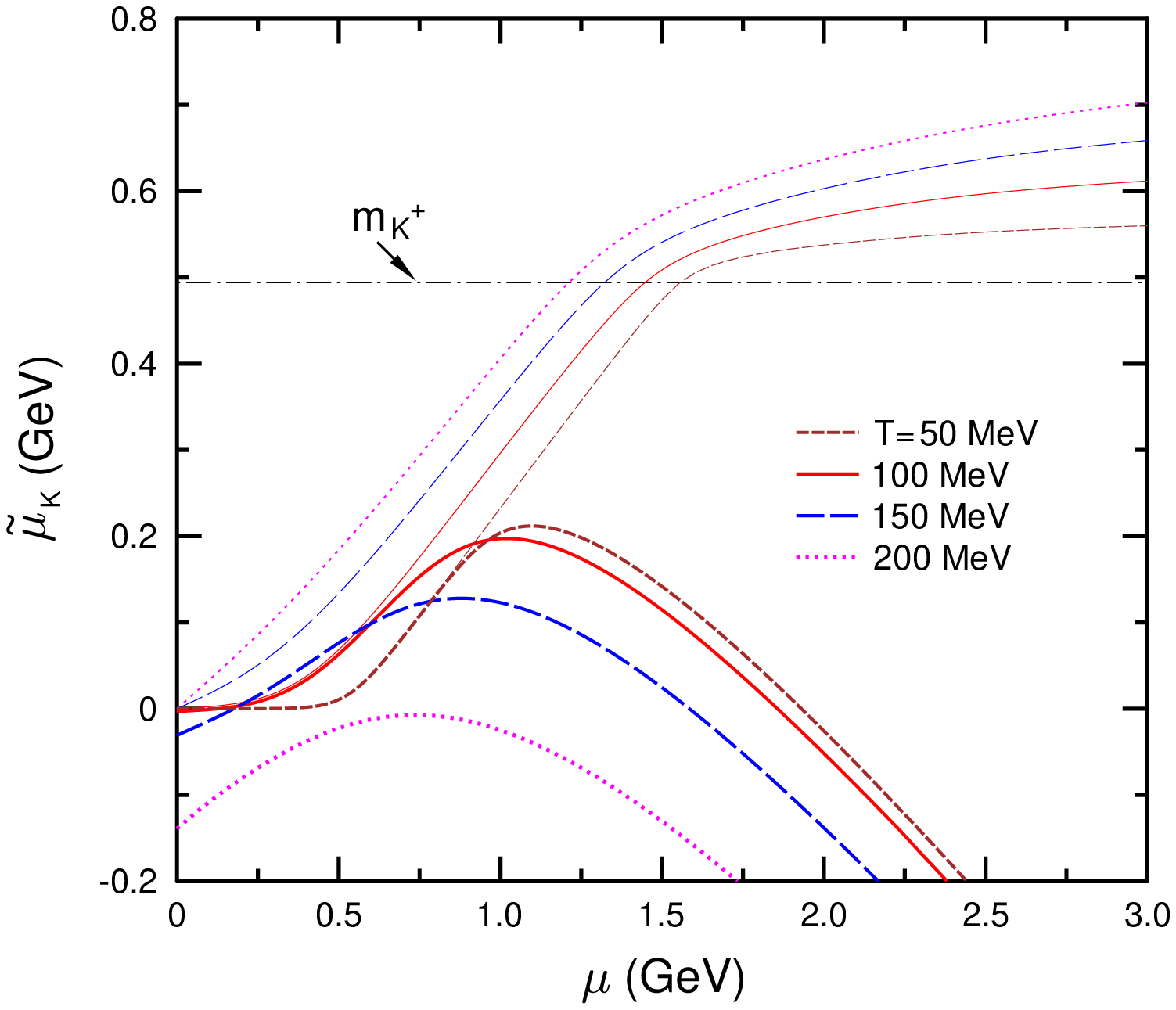}}
\caption{
The effective chemical potential of kaons as a function of $\mu$ at different
tem\-peratures $T$.
Thin and thick lines are calculated assuming $v=0$ and $v=1$ fm$^3$, respectively.
The dashed--dot\-ted line shows the threshold of the $K^+$--meson condensation.
}
\label{fig:m-mk6}
\end{figure*}

In connection with nonzero $\mu_S$ one has to study the possibility of
the $K$-meson condensation in a dense hadronic medium. This problem is under
investigation for more than two decades~\cite{Kap86}.
As well known, the Bose condensation of the $i$--th mesonic species becomes possible if the
effective chemical potential (\ref{cpti}) exceeds the meson mass~$m_i$\,\footnote
{
Here we disregard possible in-medium modifications of meson properties. This question have
been studied on the mean field level in Refs.~\cite{Sch94,Sch97,Zak05}.
}.
Within the considered model one
may expect positive chemical potentials
$\widetilde{\mu}_i$ only for strange mesons with~$S_i=1$\hsp. In this respect
the most ''dangerous'' are $K^+$ mesons with mass~\mbox{$m_{K^+}\simeq 494$\,MeV}.
As one can see from Fig.~\ref{fig:m-mk6}, in the case $v=0$ the Bose--condensation
of $K^+$ mesons would be possible at $\mu\gtrsim 1$ GeV. However, in the calculation
with high enough~$v$\hsp, the effective chemical potential $\widetilde{\mu}_K$
does not exceed the condensation threshold at any $\mu$ and~$T$. Thus, the EVC result
in total suppression of the $K^+$-condensation in che\-mi\-cally equilibrated hadronic matter with
$n_S=0$\hsp. It would be interesting to repeat this analysis for nonzero $n_S$ expected
in stellar environments.

\subsection{Hadron multiplicity ratios in heavy--ion collisions}

In this section we discuss an interesting behavior of the $K^+/\pi^+$
and $\Lambda/\pi^-$ multiplicity ratios observed in relativistic
heavy--ion collisions by the NA49 Collaboration~\cite{Afa02,Alt08}.
The experimental data show a peak (''horn'') at low SPS energies.
Our consideration follows closely the thermal model used in Refs.~\cite{Cle93,Bra96}.
\begin{table}[ht!]
\caption{Temperature and chemical potentials obtained by thermal fit
of hadron ratios in central Au+Au and Pb+Pb
collisions at different c.m. bombarding energies $\sqrt{s_{NN}}$.}
\label{tab2}
\vspace*{3mm}
\begin{tabular}{r|r|r|r}\hline
$\sqrt{s_{NN}}$\hsp, GeV&~$T$, MeV&~$\mu$\hsp, MeV&~$\mu_S$\hsp, MeV\\
\hline
 2.70~& 67~&  735~& 122.5\\
 3.32~& 85~&  668~& 115.7\\
 3.84~& 97~&  621~& 111.7\\
 4.30~& 106~& 584~& 108.5\\
 4.85~& 117~& 545~& 107.6\\
 6.41~& 132~& 460~& 94.9\\
 7.74~& 141~& 405~& 86.4\\
 8.87~& 147~& 368~& 80.9\\
 12.4~& 155~& 287~& 64.0\\
 17.3~& 159~& 219~& 48.5\\
 62.4~& 164~&  69~& 15.1\\
  130~& 164~&  34~&  7.4\\
  200~& 164~&  22~&  4.8\\
 5500~& 164~& 0.8~&  0.2\\
 \hline
\end{tabular}
\end{table}
It is assumed that production of se\-con\-dary hadrons in a nuclear collision
can be described as the emission from a statistically equilibrated
volume of hadronic matter characterized by certain temperature and baryon chemical potential.
Within this model the kaon to pion multiplicity ratio equals $n_K^*/n_\pi^*$
where $n_K^*$ and~$n_\pi^*$ are, respectively, the equilibrium densities of kaons and pions
including those produced in decays of hadronic resonances.
We apply the following relations
\bel{pikden}
n_\pi^*=\sum\limits_i d_i^\pi n_i,~~~n_K^*=\sum\limits_i d_i^K n_i\hsp.
\ee
Here \mbox{$n_i=n_i\hsp (\mu,\mu_S, T)$} is the partial density of $i$-th hadrons at fixed
$\mu,\mu_S$ and $T$ calculated by~\re{pard}.
As before, $\mu_S$ is determined from the condition of strangeness neutrality~(\ref{strd}).
The numerical values of $d_i^\pi$ and $d_i^K$ are listed in Table~\ref{tab1}.
By definition, these factors equal unity for directly produced pions and kaons, and they
are zero for stable hadrons with~\mbox{$i\neq\pi,K$}.
In our calculations we use the chemical freeze-out parameters $\mu,T$ determined from thermal
fits of hadronic ratios
measured at the AGS, SPS and RHIC energies. Namely, we apply the parametrizations of $\mu,T$
as functions of the bombarding energy $\sqrt{s_{NN}}$ given in Ref.~\cite{And08}.
In this way we get the values listed in Table~\ref{tab2}. The last column of the table
gives $\mu_S=\mu_S\hsp (\mu,T)$ calculated by using Eqs.~(\ref{pres})--(\ref{cceq}),
(\ref{strd})--(\ref{strd1}) for the case $v=1$\,fm$^3$.
The last line is obtained by extrapolating the parametrizations of $\mu,T$
to the LHC energy.

\begin{figure*}[htb!]
\centerline{\includegraphics[width=0.7\textwidth]{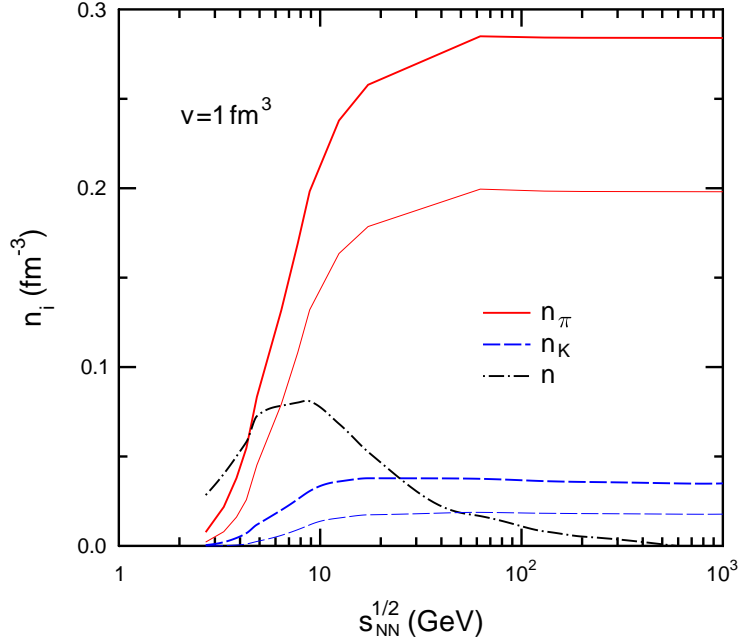}}
\vspace*{-4mm}
\caption{
Total densities of pions (thick solid line) and kaons (thick dashed line)
at chemical freeze--out in central heavy-ion collisions as functions of c.m.
bombarding energy ($v=1$\,fm$^3$). Thin lines give the contributions to pion
and kaon densities from resonance decays. The dashed-dotted line shows the net baryon density.
}
\label{fig:s-n7}
\end{figure*}
\begin{figure*}[htb!]
\centerline{\includegraphics[width=0.7\textwidth]{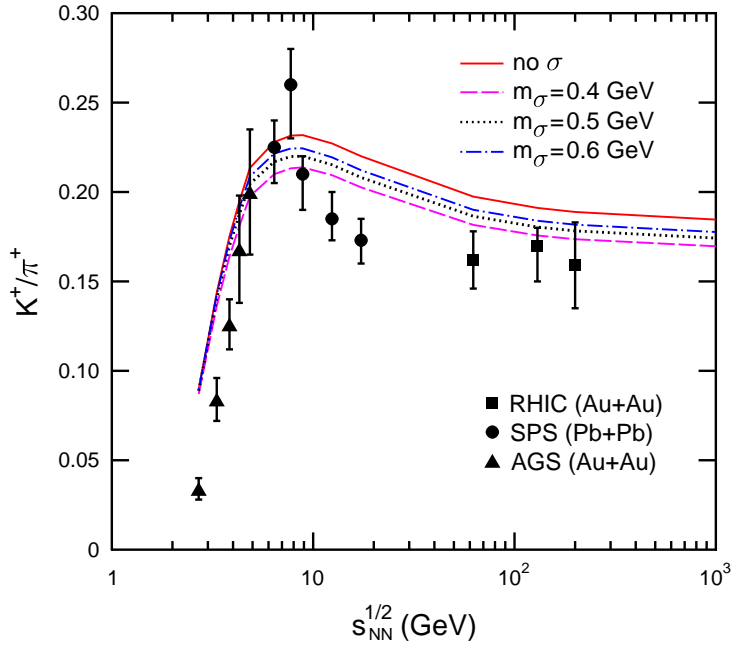}}
\vspace*{-4mm}
\caption{
The $K^+/\pi^+$ multiplicity ratio in central nuclear collisions as a function
of c.m. bombarding energy ($v=1$\,fm$^3$). Experimental data are taken from
Refs.~\cite{Afa02,Ahl00,Alt08,Abe08}.
}
\label{fig:s-kp8}
\end{figure*}
In Fig.~\ref{fig:s-n7} we show the densities $n_\pi^*,n_K^*$ as well as the baryon density
(at freeze-out stage), calculated by using the~$T,\mu,\mu_S$ values from Table~\ref{tab2}. According to
this calculation, the maximal baryon density of about $0.5\hsp n_0$
is reached at $\sqrt{s_{NN}}\simeq 9$\,GeV ($E_{\rm lab}=40$\,AGeV).
One can see that the pion density saturates at $\sqrt{s_{NN}}\gtrsim 40$\,GeV
on the level of about~0.3~fm$^{-3}$. On~the other hand, the maximal kaon
density $\sim 0.04$~fm$^{-3}$ is reached already at \mbox{$\sqrt{s_{NN}}\sim 10$\,GeV}.
It is interesting to note  that at high energies the contributions of resonance decays
reach about~70\% and 50\% in the case of pions and kaons, respectively.

In Fig.~\ref{fig:s-kp8} we
compare our results for the $K^+/\pi^+$ ratio\,\footnote
{
Using the isospin symmetry we write $K^+/\pi^+=1.5\hsp n_K^*\hspace*{-1pt}/n_\pi^*$\hsp.
}
with experimental data obtained at
the AGS~\cite{Ahl00}, SPS~\cite{Afa02,Alt08} and RHIC~\cite{Abe08} energies.
The solid line represents our standard calculation, where the broad
$f_0(600)$ resonance ($\sigma$ meson) is disregarded. One can see that
our model is able to reproduce qualitatively the observed $K/\pi$ ratios but the agreement with
data is not perfect. In particular, the experimental data exhibit a much sharper
peak at $\sqrt{s_{NN}}\simeq 7.74$\,GeV. Choosing different values of $v$ gives only slight
modifications and does not improve the shape of the $K/\pi$ excitation function.
This is easy to understand since the EVC are practically cancelled out in hadron
multiplicity ratios.

\begin{figure*}[htb!]
\centerline{\includegraphics[width=0.7\textwidth]{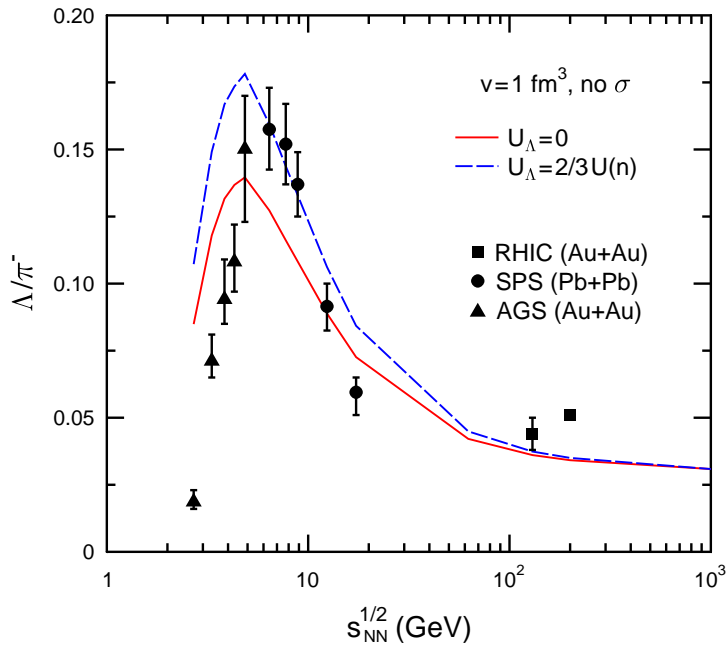}}
\vspace*{-4mm}
\caption{
The $\Lambda/\pi^-$ multiplicity ratio in central nuclear collisions as a function
of c.m. bombarding energy ($v=1$\,fm$^3$). The dashed lines are calculated
with inclusion of additional mean-field potential of $\Lambda$ hyperons.
Experimental data are taken from
Refs.~\cite{Afa02,Ahl00,Alt08,Abe08,Pink02,Alb02}.
}
\label{fig:s-lp9}
\end{figure*}
Recently, the authors of Ref.~\cite{And08} have noticed that
these data can be better reproduced by the thermal model which takes into account
the contribution of the $f_0(600)$ resonance. To verify this observation,
we have made additional calculations including this meson. These calculations are done
in the zero-width approximation assuming different masses of $\sigma$ meson,~$m_\sigma$,
and using the values \mbox{$g_\sigma=1, d_\sigma^{\hsp\pi}=2, d_\sigma^K=0$}. As seen
in Fig.~\ref{fig:s-kp8} visible changes in the~$K^+/\pi^+$ ratio take place at
$\sqrt{s_{NN}}\gtrsim 5$\,GeV. However, even at smallest~\mbox{$m_\sigma\simeq 0.4$ GeV},
due to increased pion multiplicity, this ratio drops only by about 10\%. At the same time,
the observed values in the peak region are underestimated stronger in this case. As expected,
the shifts of the $K^+/\pi^+$ ratio become smaller for larger $m_\sigma$\hsp.
On the basis of this analysis, we conclude that the inclusion of the $\sigma$ meson
does not solve the problem. According to Ref.~\cite{Bec04}, to obtain a better fit
of data one should introduce an additional parameter $\gamma_S<1$ which is
responsible for the suppression of strange hadron yields compared to the
equilibrium model predictions.

We have also calculated the excitation function of the $\Lambda/\pi^-$ ratio.
The results are shown in Fig.~\ref{fig:s-lp9}. One can see that the model can
qualitatively reproduce the experimental data, although the peak height
is somewhat underestimated. Our analysis shows that the agreement with the data can be
improved if one takes into account an additional mean--field potential
of $\Lambda$ hyperons, $U_\Lambda=\frac{2}{3}\hsp U(n)$\,\,\footnote
{
The factor $2/3$ comes from a naive consideration based on the $SU(3)$ flavor symmetry.
},
where $U(n)$ is the corresponding Skyrme-like potential for nucleons introduced
in Sect.~\ref{Spar}.

Finally, we would like to emphasize that a nonmonotonous behavior
of the $K/\pi$ and~$\Lambda/\pi$ excitation functions is obtained without any
reference to the QGP. Within our model this anomaly is mainly related to
the strangeness neutrality condition which in turn requires nonzero $\mu_S$ in
the hadronic phase. In the Boltzmann approximation the $K/\pi$ and $\Lambda/\pi$ ratios
are approximately proportional to $y_K$ and $y_\Lambda$ defined as
\bel{yfd}
y_K=\textrm{exp}\left[(\mu_S-m_K+m_\pi)/\hsp T\right],~~~y_\Lambda=\textrm{exp}\left[
(\mu-\mu_S-m_\Lambda+m_\pi)/\hsp T\right].
\ee
According to our calculations
these functions have maxima at $\sqrt{s_{NN}}\simeq 9$ and~$5$\,GeV, \mbox{respectively}.
The appearance of these maxima follows from a nonmonotonous behavior~of the
strangeness fugacity $\textrm{exp}(\mu_S/T)$ along the chemical freeze-out line\,\footnote
{
As was mentioned on page~\pageref{musref}, $\mu\sim\frac{1}{2}\,\mu_S$, so that
$\mu-\mu_S\sim\mu_S$\hsp.
}.
As demonstrated above, this behavior agrees qualitatively with experimental data.
The inclusion of resonance decays makes the $K/\pi$ and~$\Lambda/\pi$ peaks
more pronounced.

\section{Quark--gluon phase within the bag model\label{QEOS}}

To calculate thermodynamic properties of a~baryon--rich QGP
we use a simple bag model with perturbative corrections of the order $\alpha_s$\hsp.
Gluons ($i=g$) and
light quarks (\mbox{$i=q,\ov{q}$}) are considered as massless point--like particles,
but for strange quarks ($i=s,\ov{s}$) we introduce a~nonzero mass $m_s$\hsp.
Nonperturbative effects are introduced via the bag constant~$B$\hsp.
In~the chemically equilibrated QGP, one can again express the chemical potential
of the $i$--th particle by~\re{cceq}, where $\mu$ and $\mu_S$ are now the baryon
and strange chemical potentials of the quark--gluon system. As a result, we get the relations
\mbox{$\mu_g=0$}\hsp, \mbox{$\mu_q=-\hspace*{.5pt}\mu_{\ov{q}}=\mu/3$}\hsp,
$\mu_s=-\mu_{\ov{s}}=\mu/3-\mu_S$\hsp.

Following Ref.~\cite{Iva05} we take into account perturbative corrections
by introducing additional constant factors $1-\xi$ and $1-0.8\hsp\xi$ into the kinetic pressure
of quarks and gluons, respectively. Here $\xi\sim\hsp\alpha_s$ is the model parameter
which we fix by the comparison with lattice data. Within such an approach one gets the following
expression for pressure of the QGP:
\begin{eqnarray}
&&\hspace*{-5mm}P=\left(\widetilde{N}_g+\frac{21}{2}\widetilde{N}_f\right)\frac{\pi^2T^4}{90}
+\widetilde{N}_f\left(\frac{\mu^2T^2}{18}+\frac{\mu^4}{324\hsp\pi^2}\right)
+\frac{1-\xi}{\pi^2}\int\limits_{m_s}^\infty d\epsilon\hsp (\epsilon^2-m_s)^{3/2}
\left\{\frac{1}{e^{\frac{\epsilon-\mu_s}{T}}+1}\right.\nonumber\\
&&\left.+\frac{1}{e^{\frac{\epsilon+\mu_s}{T}}+1}\right\}-B\,.
\label{preq}
\end{eqnarray}
Here $\widetilde{N}_g=16\hsp(1-0.8\hsp\xi)$ is the effective number of gluons
and $\widetilde{N}_f=2\hsp (1-\xi)$ is the
effective number of light flavors. The third term in \re{preq} gives the contribution
of strange quarks and antiquarks. Corresponding expressions for densities of baryon charge
$n$, strangeness $n_S$, energy $\epsilon$ and entropy $s$ can be obtained from~\re{preq}
by using the thermodynamic relations~(\ref{ther}), (\ref{gdeq}). One can easily see that the
strangeness density of the QGP, \mbox{$n_S=n_{\ov{s}}-n_s=\partial P/\partial\mu_S$}\hsp, vanishes
if $\mu_s=\mu_S-\mu/3=0$\,.

\begin{figure*}[hbt!]
\centerline{\includegraphics[width=0.7\textwidth]{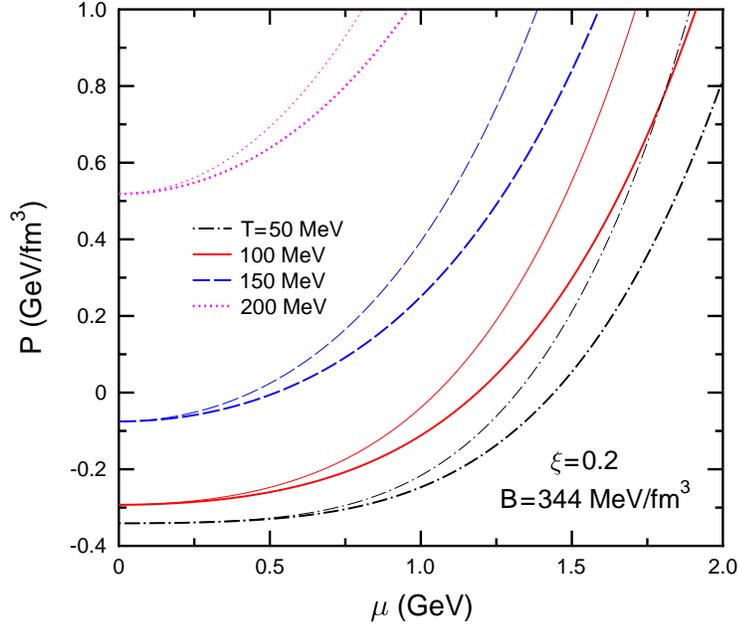}}
\caption{
Same as Fig.~\ref{fig:m-p4}, but for pressure isotherms of the quark--gluon plasma.
Notice the appearance of negative pressure values at low $T$ and $\mu$\hsp.
}
\label{fig:m-p10}
\end{figure*}

Below we fix $m_s=150$\,MeV and take the same value $B=344$\,MeV/fm$^3$
as in our previous fluid--dynamical calculations (EOS--I)~\cite{Sat07,Sat07b} of heavy-ion
collisions at RHIC energies. We have calculated the thermodynamic functions of the QGP
for different values of the parameter $\xi$\hsp.
Motivated by the comparison with lattice data
(see Fig.~\ref{fig:t-e11}), we further use the value $\xi=0.2$ in our calculations.
The advantage of the bag model is that it can be used in the region of nonzero chemical
potential, which is still not accessible by lattice calculations.
In Fig.~\ref{fig:m-p10} we show pressure isotherms of strange (\mbox{$\mu_S=0$}) and
non\-strange~($\mu_S=\mu/3$) QGP. Similarly to the hadronic matter (see Fig.~\ref{fig:m-p4}),
in the case of nonzero~$\mu_S$ the pressure is reduced compared to the calculation with $\mu_S=0$\hsp.

\section{The deconfinement phase transition and properties of mixed phase}

\subsection{Phase equilibrium conditions\label{PEC}}

Below we use our phenomenological approach, where the hadronic and quark-gluon phases are described
by two different models, to study the possibility of a first order deconfinement phase
transition. If such a transition exists, then only three types of
equilibrium states are possible at a given point in the ($\mu,\mu_S,T$)
space: the hadronic phase (HP), the quark-gluon phase (QP) and the mixed phase (MP).
Outside the MP region the
stable phase is the one which has a higher pressure~\cite{Lan80}.
The Gibbs condition of equilibrium between the domains of different phases in the MP can be written
as
\bel{gibbs}
P_H\hsp(\mu,\mu_S,T)=P_Q\hsp(\mu,\mu_S,T)\,.
\ee
Here and below we mark thermodynamic functions of the hadronic and quark--gluon phases
by indices $H$ and $Q$\hsp, respectively. The expressions for pressure $P_H$ and $P_Q$
are given in Sect.~\ref{HEOS} and \ref{QEOS}.

In a baryon--rich mixed phase the condition of zero total strangeness leads to
the strangeness--antistrangeness separation phenome\-non~\cite{Gre87}.
At nonzero $\mu$ and $T$, the equations $n_{SH}=n_{SQ}=0$ may hold
simultaneously with \re{gibbs} only in a single point of the~($\mu,\mu_S,T$) space.
But in general,
the strangeness numbers of coexisting domains are nonzero and
compensate each other only on average
over the total volume of the system. Let $\lambda$ denotes the volume fraction of
hadronic domains in the MP:
\bel{vfrac}
\lambda=\frac{V_H}{V_H+V_Q}\in [0;1]\hsp,
\ee
where $V_H$ and $V_Q$ are  the total spatial volumes occupied by hadrons and the QGP,
respectively. Then the condition of zero total strangeness may be written as
\bel{sden}
n_S=\lambda\hspace*{1.5pt}n_{SH}+(1-\lambda)\hspace*{1.5pt}n_{SQ}=0\hsp.
\ee
This condition holds if $n_{SH}$ and $n_{SQ}$ have different signs. At fixed $\mu,\lambda$
the phase transition temperature $T_c$ is determined by solving Eqs.~(\ref{gibbs}), (\ref{sden}).
The phase transition region in the~$\mu-T$ plane is not a line,
but a strip~\cite{Gor05}\, $T=T_c\hsp(\mu,\lambda)$ where $T_c$ is decreasing function of~$\lambda$\,.
The lines $T=T_c\hsp(\mu,1)$ and $T=T_c\hsp(\mu,0)$ give the hadronic and quark--gluon boundaries of
the~MP. Our calculations show (see Fig.~\ref{fig:m-t15}) that the strip's width decreases with
increasing excluded volume $v$\hsp.

The energy, baryon and entropy densities in the MP are calculated by using the first equality
of (\ref{sden}) with the replacement of $n_S$ by $\epsilon, n$ and $s$\hsp, respectively.
At given $\mu,T$ we use the values of $\mu_S,\lambda$ obtained by solving
Eqs.~(\ref{gibbs}), (\ref{sden})\hsp.

\subsection{Results for baryon-free matter\label{BFM}}

\begin{figure*}[hbt!]
\centerline{\includegraphics[width=0.7\textwidth]{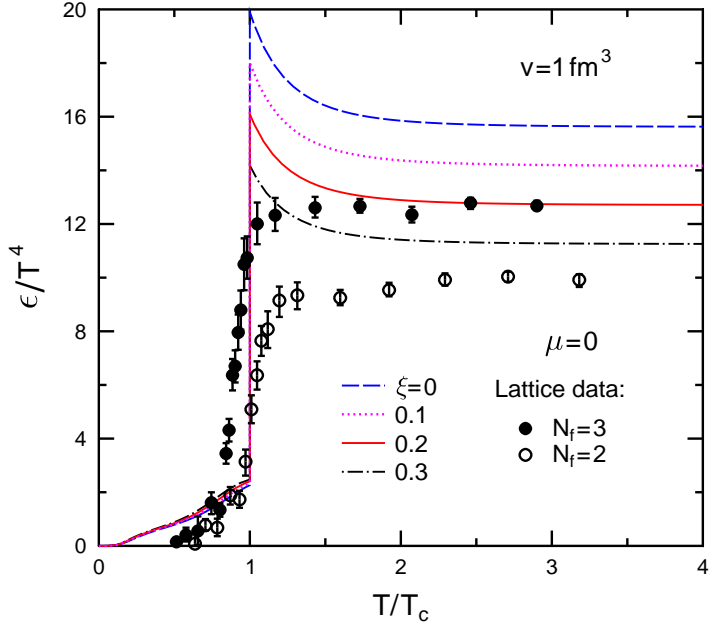}}
\vspace*{-4mm}
\caption{
Scaled energy density of baryon--free matter as a function of
scaled temperature at different values of $\xi$. All results are obtained
with $v=1\,\textrm{fm}^3$ and $B=344\,\textrm{MeV/fm}^3$.
$T_c$~is the critical temperature of the deconfinement phase transition (see
Table~\ref{tab3})\hsp.
Open and closed circles show the lattice data~\cite{Kar01} with the number of
quark flavors $N_f=2$ and~3\hsp, respectively.
}
\label{fig:t-e11}
\end{figure*}
In this section we present our results for the
baryon--free system ($\mu=\mu_S=0$), which can be compared with lattice
calculations. Table~\ref{tab3} gives critical temperatures $T_c$ obtained for different
values of $\xi$\hsp.
\begin{table}[h!]
\caption{Critical temperature of the deconfinement phase transition
in the baryon--free matter ($v=1\,\textrm{fm}^3, B=344\,\textrm{MeV/fm}^3$)\hsp.}
\label{tab3}
\vspace*{3mm}
\begin{tabular}{c|c|c|c|c}\hline
$\xi$&0&0.1&0.2&0.3\\
\hline
$T_c$\hsp, MeV&~155.4&~159.9&~165.1&~171.3\\
\hline
\end{tabular}
\end{table}
In Fig.~\ref{fig:t-e11} we compare the model predictions for $\epsilon/T^4$
with the lattice calculations from Ref.~\cite{Kar01}.
One can clearly see a pronounced peak just above $T_c$ which
is not present in the lattice data. This is a well--known artefact of the
bag model (see e.g.~\cite{Sat07b})\hsp. At high temperatures
a good agreement with the lattice data can be achieved by
choosing~$\xi\simeq 0.2$\,.
Based on these results, we use $\xi=0.2$ in the following calculations.
\begin{figure*}[hbt!]
\vspace*{-1mm}
\centerline{\includegraphics[width=0.7\textwidth]{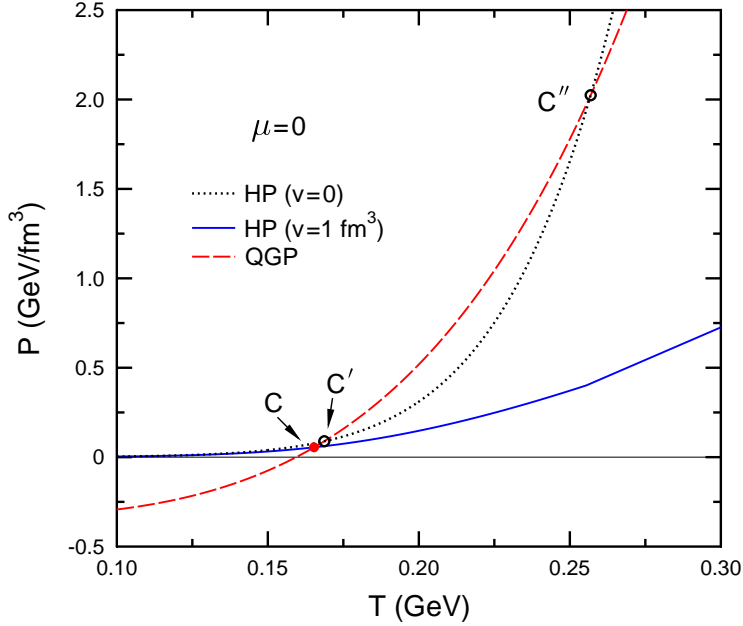}}
\vspace*{-4mm}
\caption{
Pressure of baryon--free matter as a function of temperature.
The solid and dotted lines show pressure of the HP
calculated with $v=1$\,fm$^{3}$ and $v=0$\hsp, respectively.
The dashed line shows pressure of the QGP calculated with
$B=344\,\textrm{MeV/fm}^3$ and $\xi=0.2$\hsp.
}
\label{fig:t-p12}
\end{figure*}

\begin{figure*}[htb!]
\centerline{\includegraphics[width=0.7\textwidth]{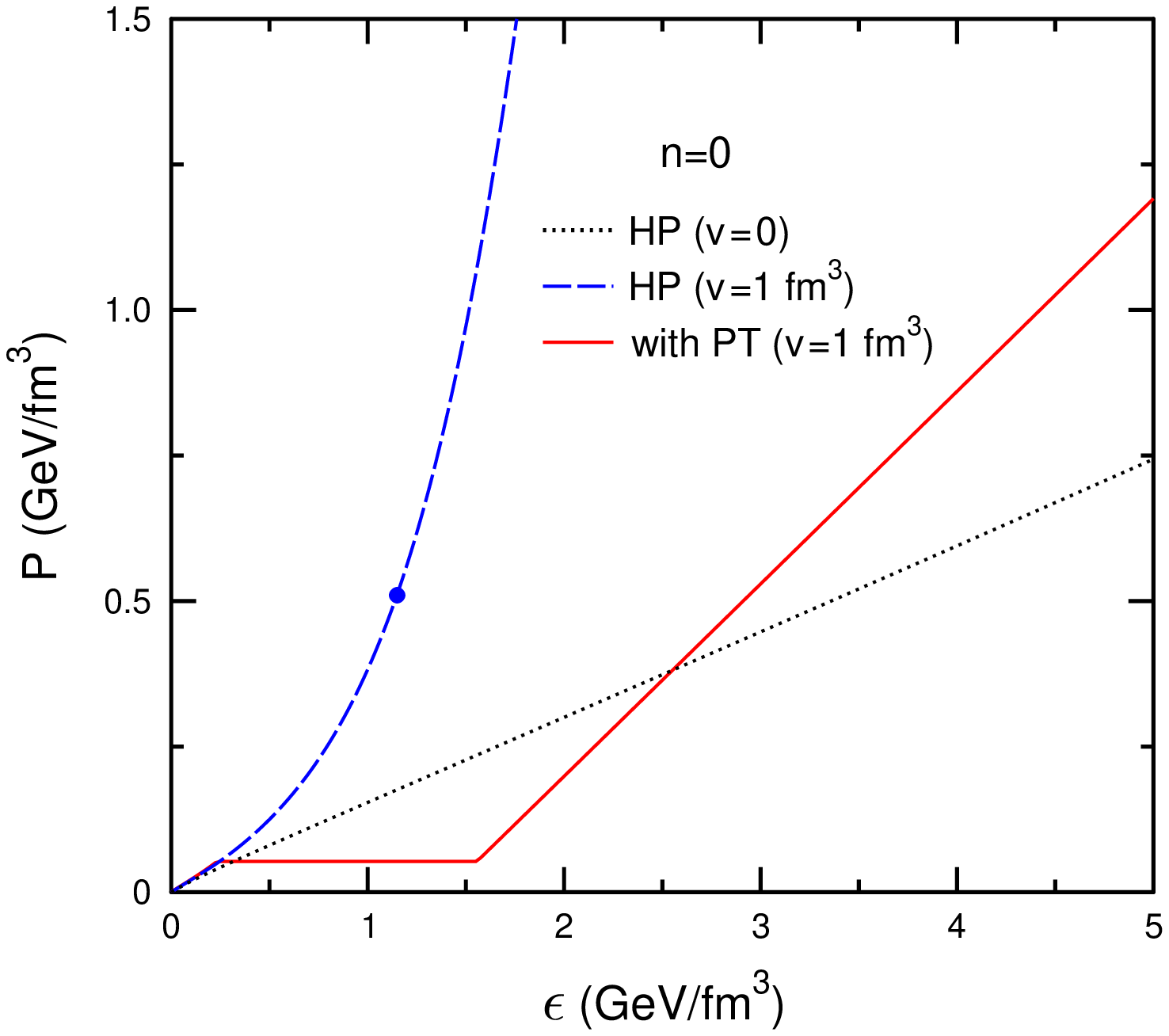}}
\caption{
Pressure of baryon--free matter as a function of energy density.
Dashed and dotted lines show pressure of a hadronic system calculated with
$v=1$\,fm$^{3}$ and $v=0$, respectively. The solid line
shows pressure calculated with inclusion of EVC and phase transi\-tion
effects. Above the dot the~HP has superluminal sound
velocities.
}
\label{fig:e-p13}
\end{figure*}
\begin{figure*}[htb!]
\centerline{\includegraphics[width=0.7\textwidth]{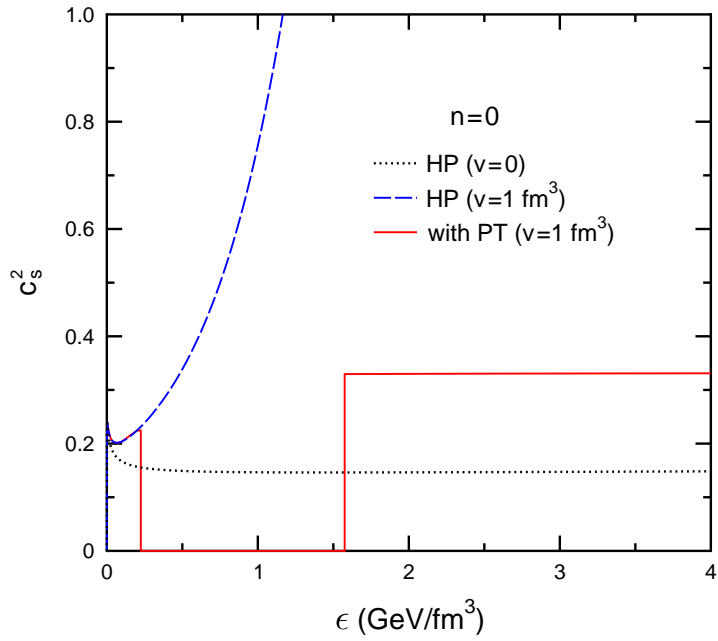}}
\caption{
Same as Fig.~\ref{fig:e-p13}, but for sound velocity squared.
}
\label{fig:e-cs14}
\end{figure*}

A more detailed information about the EOS of baryon--free systems is given in
Figs.~\mbox{\ref{fig:t-p12}--\ref{fig:e-cs14}}\hsp. Figure~\ref{fig:t-p12} represents the
graphic solutions
of~\re{gibbs} for zero and nonzero $v$\hsp. In the case~$v=1$~fm$^3$
the curves $P_H(T)$ and $P_Q(T)$ intersect in a single critical
point~$C$\hsp. On the other hand, in the calculation with $v=0$ two crossing points
$C^{\hspace*{0.7pt}\prime}$ and $C^{\hsp\prime\prime}$ are present.
The second point corresponds to the transition QGP $\to$ HP at the temperature
\mbox{$T=T_{C^{\hsp\prime\prime}}>T_{C^{\hsp\prime}}$}.
Such unusual behavior, not supported by lattice data, is explained by a too steep
rise of pressure of point--like hadrons with increasing number of species
at high temperatures.

To perform fluid--dynamical modelling of heavy--ion collisions, one should know pressure
as a function of the energy density $\epsilon$ and net-baryon density $n$\,.
Such a representation of the EOS of the baryon--free matter is shown in Fig.~\ref{fig:e-p13}.
In this figure the deconfinement phase transition shows up as a horizontal line connecting the HP and QP.
This line corresponds to the MP states. The parameters of the phase transition, calculated
for $v=1$\,fm$^3$, are close to those in the equation of state
\mbox{EOS-I} used in Refs.~\cite{Sat07,Sat07b}. The dashed line shows metastable states of the
HP extended into the region of large $\epsilon$. At such energy densities, a much
harder EOS of the HP is predicted compared to the calculation with $v=0$ (shown by the dotted line).
On the other hand, both calculations give similar results at low $\epsilon$\hsp.
In fluid--dynamical simulations of heavy--ion collisions~\cite{Sat07,Sat07b,Tea01}
two scenarios, with and without the phase transitions, are often compared
to check the sensitivity of observables to the EOS.
However, from comparing the dashed and dotted lines in Fig.~\ref{fig:e-p13}, it is evident
that such analysis could be misleading if the role of repulsion in a
dense hadronic system is ignored.

The sound velocity $c_s$ is an important characteristic of the EOS which gives
the speed of small perturbations of matter in its local rest frame.
Within the ideal hydrodynamics the sound velocity squared is equal to~\cite{Lan80}
\bel{cs2r}
c_s^{\hsp 2}=\left(\frac{\ds\partial P}{\ds\partial\epsilon}\right)_{\hspace*{-2pt}\sigma}=
\left(\frac{\ds\partial P}{\ds\partial\epsilon}\right)_{\hspace*{-2pt}n}
+\frac{n}{w}\left(\frac{\ds\partial P}{\ds\partial n}\right)_{\hspace*{-2pt}\epsilon},
\ee
where $\sigma=s/n$ is the entropy per baryon and $w=\epsilon+P$ is the enthalpy density.
In the second equality we have applied the thermodynamic relation ($n_S=0$)\hsp:
\bel{thr1}
d\sigma=\frac{1}{nT}\hspace*{1.5pt}(\hsp d\epsilon-\frac{w}{n}\hspace*{1.5pt} dn)\,.
\ee
In the case of baryon--free matter ($n=0$) $c_s^{\hsp 2}$ is equal to the slope of pressure
as a function of $\epsilon$\hsp.

The causality condition $c_s<1$ should be fulfilled in any model
consistent with relativistic kinematics.
However, a simple Van der Waals approach does not guarantee this property,
as has been already mentioned in Refs.~\cite{Ris91,Ven92}.
In a gas of hard spheres the acausal behavior is associated with an
implicit assumption that the signal propagates instantaneously over the sphere extension.
As will be shown below, the causality condition $c_s<1$ is indeed violated
in the HP, but  only at rather high baryon densities and small temperatures.
For illustration, Fig.~\ref{fig:e-cs14} shows $c_s^2$ for baryon--free matter.
One can see that~$c_s^{\hsp 2}$ is close to 1/3  in the QP, vanishes in the~MP and equals
$0.2-0.25$ (at $v=1$ fm$^3$) in the~HP. The values $c_s>1$ are reached only
in metastable hadronic states with $\epsilon\gtrsim 1$~GeV/fm$^3$ which are not
realized in the equilibrated matter.

\subsection{Phase diagram in $\mu-T$ plane}

\begin{figure*}[htb!]
\centerline{\includegraphics[width=0.7\textwidth]{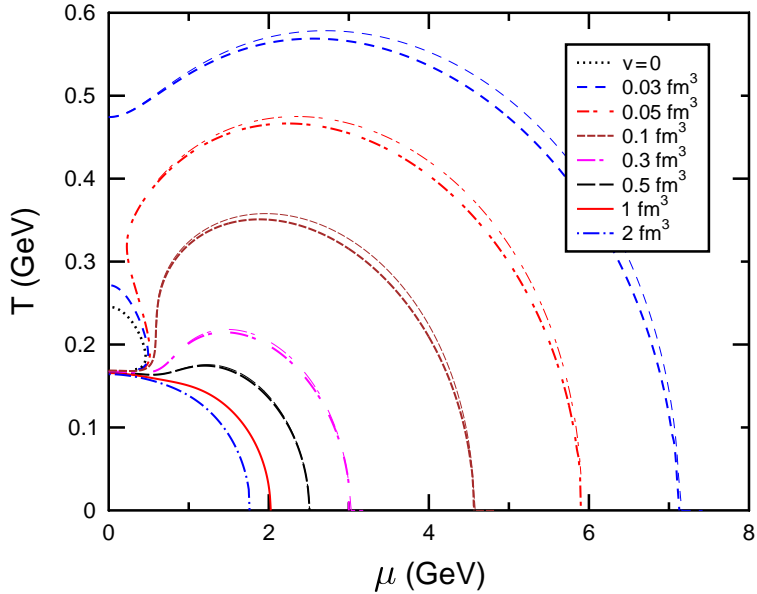}}
\caption{
Phase diagrams in the $\mu-T$ plane calculated for different $v$\hsp. Thick and thin lines
correspond to hadronic and quark--gluon boundaries of the MP, respectively.
}
\label{fig:m-t15}
\end{figure*}
Using the Gibbs condition~(\ref{gibbs}) we have calculated
the phase diagram of matter in the~\mbox{$\mu-T$} plane.
The results for different values of $v$ are given in Fig.~\ref{fig:m-t15}.
For each~$v$ we show
two boundaries of the MP corresponding to the conditions $n_{SH}=0\,(\lambda=1)$
or~$n_{SQ}=0\,(\lambda=0)$. One can see that the width of the MP region decreases with
increasing~$v$. The two boundaries are practically undistinguishable at~$v\gtrsim 0.5$ fm$^3$.
With decreasing $v$ the HP occupies larger and larger domain of the~$\mu-T$ plane. At
$v\lesssim 0.1$ fm$^3$ the phase boundary exhibits a back bending at small chemical
potentials, \mbox{$\mu<\mu_*\simeq 0.8$ GeV}. This means that at fixed $\mu<\mu_*$
and increasing $T$ three phase transitions appear: the first and third ones
from the HP to the QGP, and the second intermediate transition, from the QGP to the HP.
The third point goes to infinity at $v\to 0$\hsp.
This means that the HP is thermodynamically more stable
than the QP at asymptotically high temperatures.
At $v=0$ only a small domain of the QGP (its boundary is shown by the dotted line) remains
in the~$\mu-T$ plane. This is certainly an unphysical behavior, which clearly shows
inconsistency of the hadron resonance gas model with point-like hadrons.

\begin{figure*}[htb!]
\centerline{\includegraphics[width=0.7\textwidth]{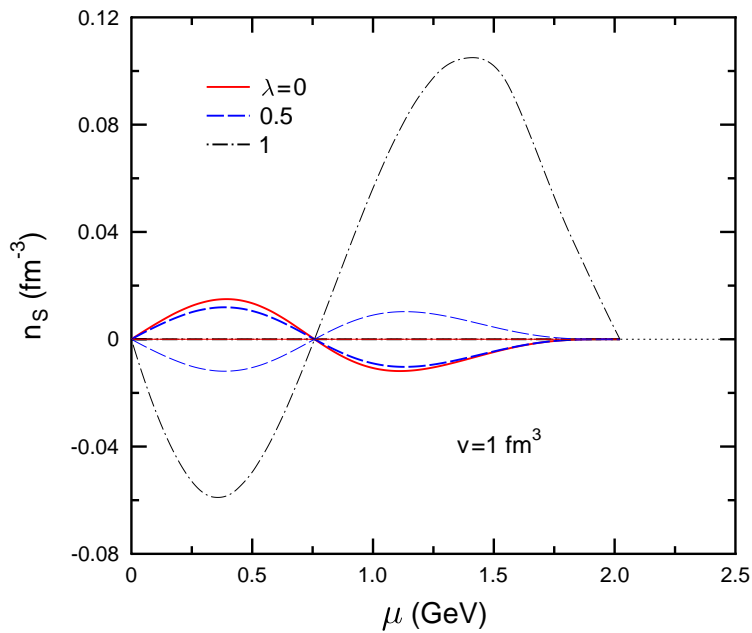}}
\vspace*{-5mm}
\caption{
Strangeness densities of the hadronic (thick lines) and quark--gluon (thin lines)
components in the MP region of the phase diagram as functions of the baryon
chemical potential. The results are shown for several values of the volume
fraction $\lambda$\hsp. For $\lambda=0$ (no hadrons) $n_{SQ}=0$\hsp.
}
\label{fig:m-ns16}
\end{figure*}
\begin{figure*}[h!]
\vspace*{-3mm}
\centerline{\includegraphics[width=0.7\textwidth]{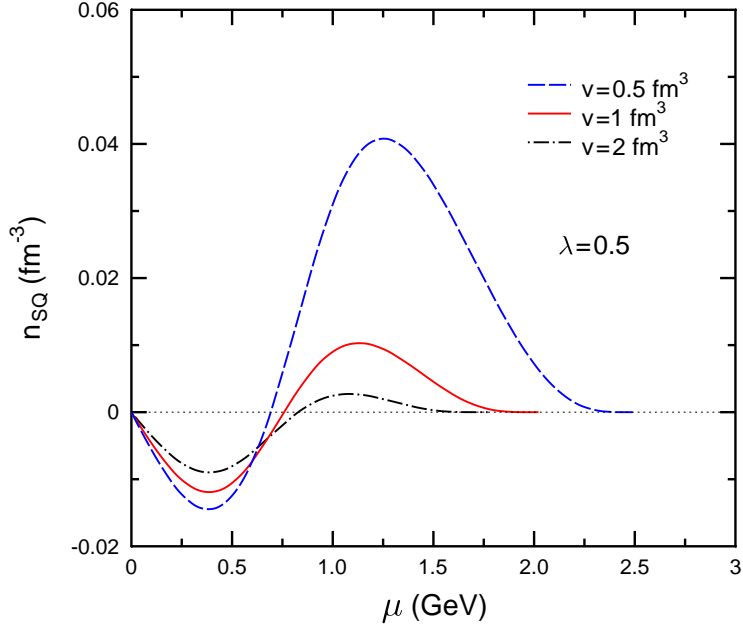}}
\vspace*{-5mm}
\caption{
Strangeness density of the quark--gluon domains in the MP ($\lambda=0.5$) at different values
of the excluded volume $v$\hsp.}
\label{fig:m-ns17}
\end{figure*}
Figure \ref{fig:m-ns16} shows the strangeness number densities $n_{SH}$ and $n_{SQ}$
across the MP (\mbox{$v=1$\,fm$^3$})\hsp.
One can see that indeed, $n_{SH}$ and $n_{SQ}$ have different signs~(see Sect.~\ref{PEC}), but their
absolute values do not exceed
$0.1$ fm$^{-3}$. At $\mu\gtrsim 0.8$ GeV (anti)\hsp strangeness in hadronic domains
is carried mostly by hyperons. According to Fig.~\ref{fig:m-ns17}, at
large chemical potentials the strangeness density $n_{SQ}$ is rather sensitive to $v$\hsp.

\subsection{Adiabatic trajectories}
\begin{figure*}[htb!]
\centerline{\includegraphics[width=0.7\textwidth]{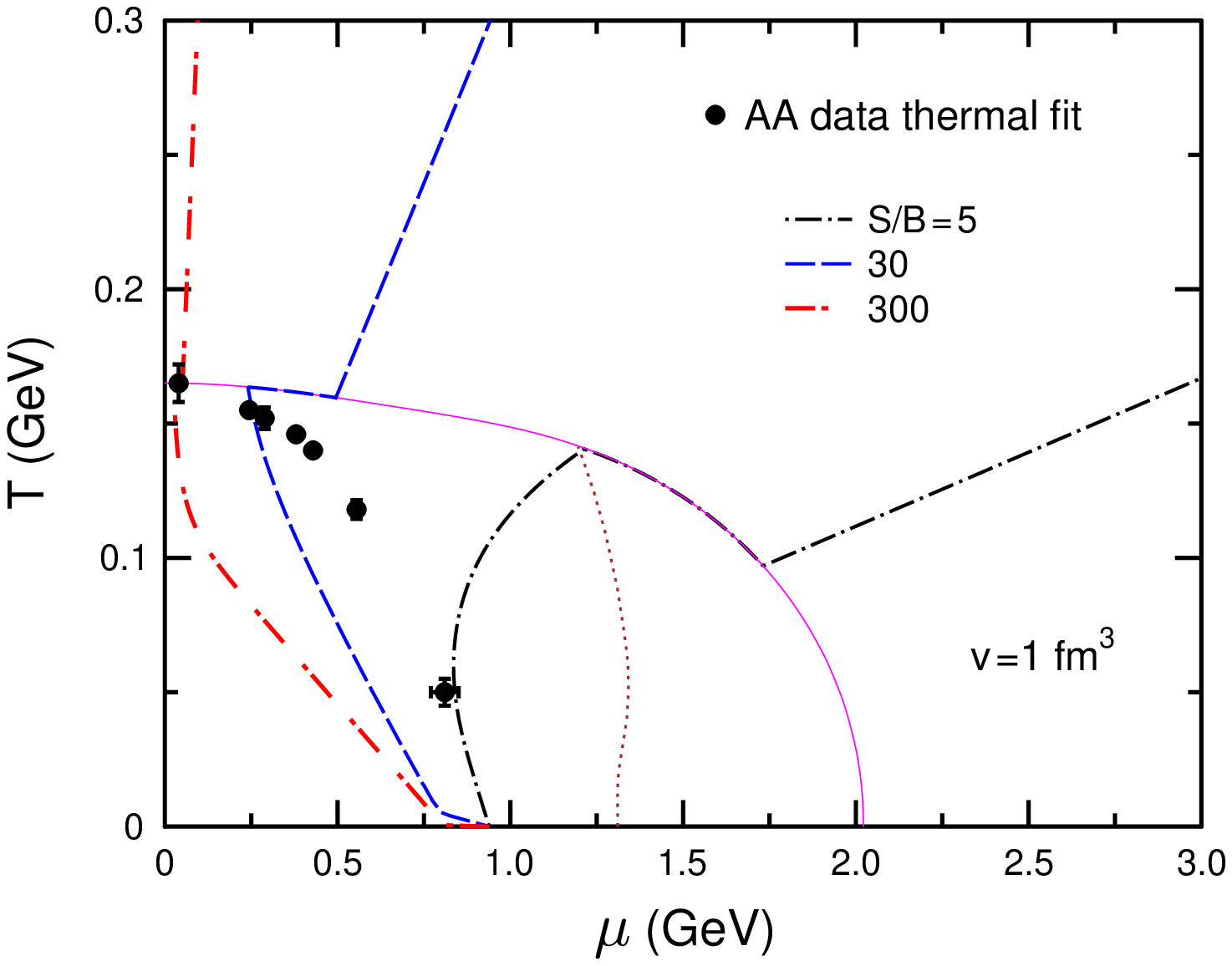}}
\caption{
Lines of constant entropy per baryon in the \mbox{$\mu-T$} plane (\mbox{$v=1$ fm$^3$})\hsp.
The solid curve represents the phase transition line for
$\lambda=0$\hsp. Full dots corresponds to the $\mu, T$ values obtained
from thermal fits of hadron yields~\mbox{\cite{And06}} observed in central Au+Au and Pb+Pb collisions
at different bombarding energies. The region
between the dotted and thin solid lines contains states with~\mbox{$c_s>1$}\hsp.
}
\label{fig:m-t18}
\end{figure*}
\begin{figure*}[htb!]
\centerline{\includegraphics[width=0.7\textwidth]{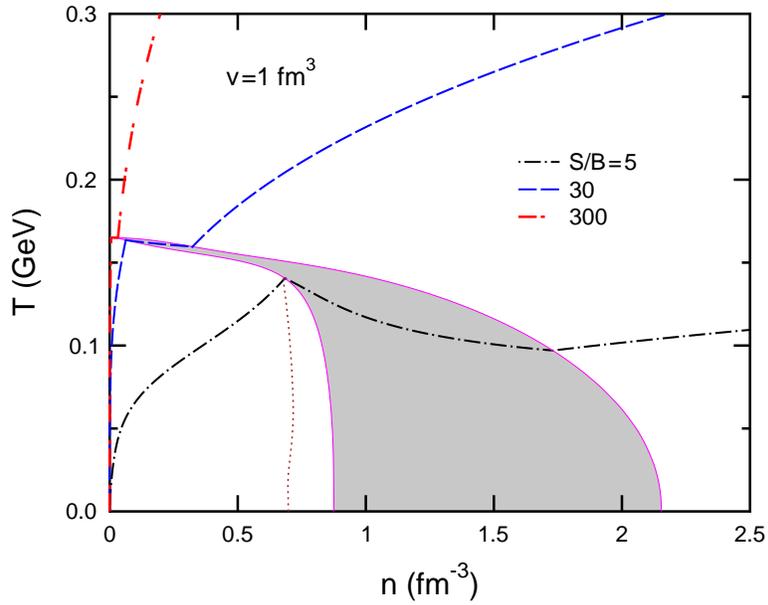}}
\caption{
Boundaries of different phases and adiabatic trajectories in the $n-T$ plane
(\mbox{$v=1$ fm$^3$})\hsp. The shaded area shows the mixed phase region.
The hadronic states on the right from the dotted line have
sound velocities $c_s>1$.
}
\label{fig:n-t19}
\end{figure*}

\begin{figure*}[htb!]
\centerline{\includegraphics[width=0.7\textwidth]{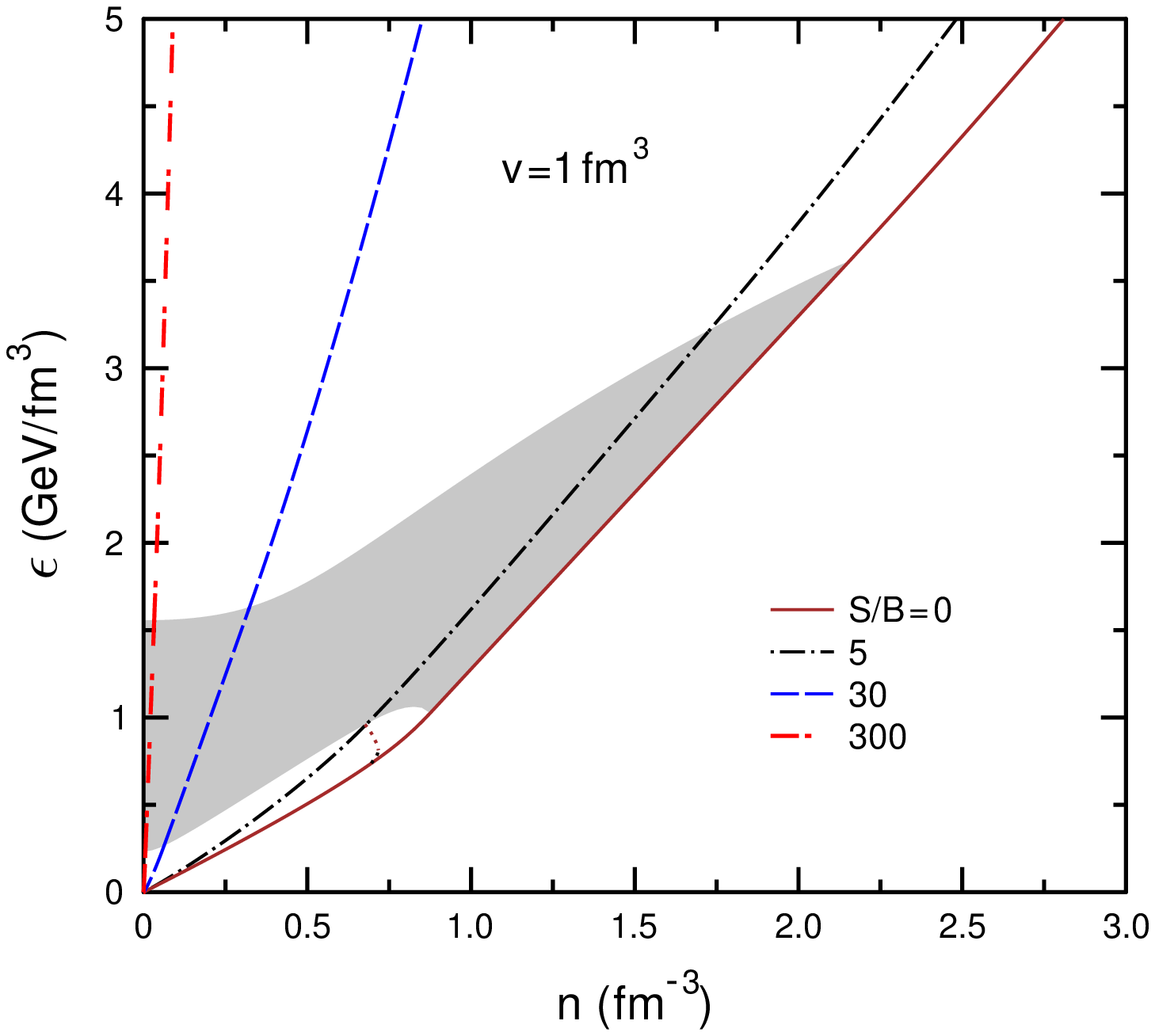}}
\caption{
Same as Fig.~\ref{fig:n-t19} but in the $n-\epsilon$ plane.
}
\label{fig:n-e20}
\end{figure*}
\begin{figure*}[htb!]
\centerline{\includegraphics[width=0.7\textwidth]{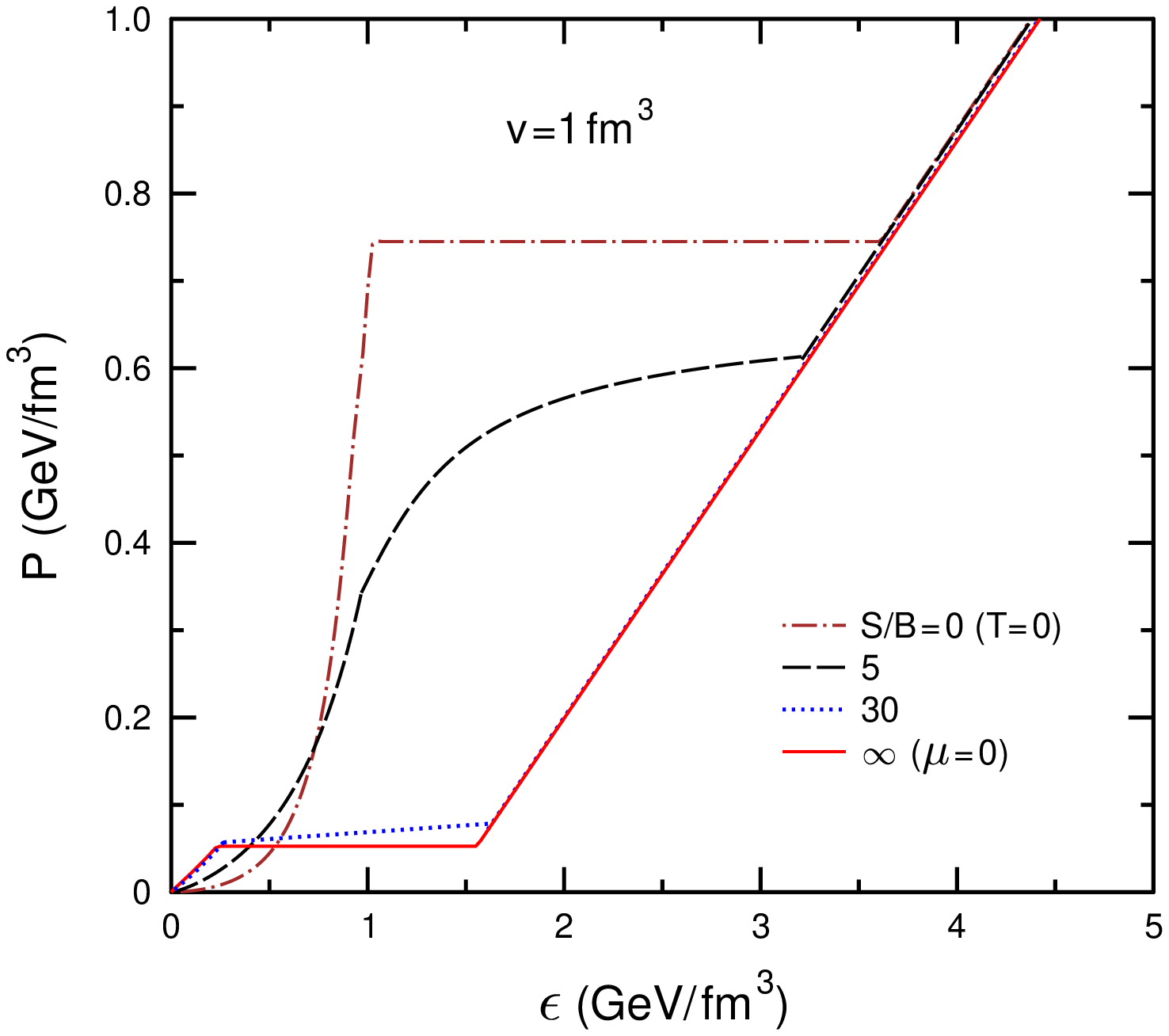}}
\caption{
Adiabatic trajectories in the $\epsilon-P$ plane (\mbox{$v=1$ fm$^3$})\hsp.
}
\label{fig:e-p21}
\end{figure*}
\begin{figure*}[htb!]
\centerline{\includegraphics[width=0.7\textwidth]{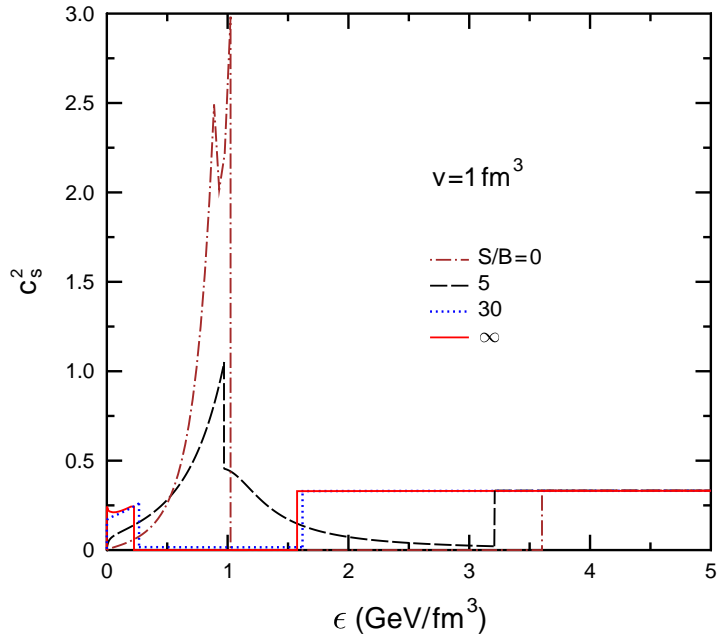}}
\caption{
Sound velocities squared in adiabatic processes with different $S/B$ as
functions of energy density (\mbox{$v=1$ fm$^3$})\hsp.
}
\label{fig:e-cs22}
\end{figure*}

Further on we consider in more details the results for $v=1$ fm$^3$.
In Figs.~\ref{fig:m-t18}--\ref{fig:e-cs22} we show the adiabatic trajectories (''adiabates'')
i.e. sets of states with equal entropy per baryon, $S/B=\sigma={\rm const}$\hsp.
As well--known~\cite{Lan87}, in the ideal hydrodynamics the entropy per baryon is conserved
in a given fluid element. Therefore, these trajectories contain an important
information about the conditions which can be realized in heavy--ion collisions.
Figure~\ref{fig:m-t18} shows
the phase diagram and corresponding adiabates in the $\mu-T$ plane.
In such a representation the adiabates with $S/B=0$ and $S/B=\infty$ are given,
respectively, by the horizontal ($T=0$) and vertical ($\mu=0$) axes.
Note that at $T\to 0$ all adiabates with finite~$\mu$ end in the point $\mu\simeq m_N$,
where $m_N=939$\,MeV is the nucleon mass. Moreover, these adiabates have a zigzag-type behavior
characteristic for a first order phase transition. This means that along the adiabatic
trajectory the temperature
grows when the system enters the coexistence region. In other words, the temperature of the
HP at $\lambda=1$ is higher than the temperature of the QP at $\lambda=0$\hsp.
Such a picture differs from the predictions of the linear $\sigma$ model
and the Nambu--Jona-Lasinio model~\cite{Sca01,Mis01} for the chiral first order
phase transition. There temperature drops in the MP. This difference may be related to the
fact that the present model includes massless gluons which are completely
ignored in Refs.~\cite{Sca01,Mis01}.

Full dots in Fig.~\ref{fig:m-t18} show states of the chemical freeze-out~\cite{Cle98} in central
collisions of heavy nuclei with c.m. bombarding energies
from GSI ($\sqrt{s_{NN}}\simeq 2.3$\,GeV) to RHIC (\mbox{$\sqrt{s_{NN}}\simeq 200$\,GeV}).
''Experimental'' values of $T$ and $\mu$ have been found~\cite{And06} from thermal fits of hadron
multiplicity ratios observed in such collisions.

A word of caution is in place here. As has been already mentioned, our model contains
superluminal sound velocities in the HP. These states are situated in
the region between the dotted and thin solid lines in Fig.~\ref{fig:m-t18}.
One can see that such states correspond to values~$S/B<5$. On the other hand,
it is known from hydrodynamical simulations
(see e.g.~\cite{Pae01}) that typical conditions realized in heavy--ion collisions
at energies \mbox{$E_{\rm lab}\gtrsim 10$ GeV} correspond to entropy per baryon $S/B\gtrsim 10$.
Therefore, an adiabatically expanding system of particles produced in ultrarelativistic
nuclear collisions does not enter the region of $c_s>1$.
However, such states can be reached in compact stars. Based
on these results, we conclude that the present model
should be considerably modified to study
properties of the HP and the deconfinement phase transition at high baryon densities
(see the discussion in Ref.~\cite{Mis02}).

In Figs.~\ref{fig:n-t19}--\ref{fig:n-e20} we show boundaries of different phases
in the $n-T$ and $n-\epsilon$ planes. Again, one can see that adiabates with $S/B\gtrsim 10$
do not enter the ''dangerous'' region with superluminal sound velocities.
The solid line in Fig.~\ref{fig:n-e20} corresponds to the limiting case of~$T=0$\hsp.
States below this line can not exist.
Figure~\ref{fig:e-p21} shows pressure adiabates versus the energy density.
According to first equality of~(\ref{cs2r}), $c_s^{\hsp 2}$\hsp--\hsp values are given by
slopes of adiabates in this representation. Note, that pressure in the MP is constant along the
lines~$\mu=\textrm{const}$ which differ from the lines $\sigma=\textrm{const}$  in the
$n-\epsilon$ plane.
The direct calculation of $c_s^2$ as a function of $\epsilon$ gives the results
shown in Fig.~\ref{fig:e-cs22}. One can see that $c_s^2$ values especially
large at zero temperature, for states near the MP boundary.
Generally, the sound velocities are nonzero in the MP,
however, they are much smaller than maximal values of $c_s$ in the HP.
The nontrivial two--peak structure of the adiabate with $S/B=0$ appears due to the contribution
of baryonic resonances ($\Delta, N^* \ldots$).
However, this behavior takes place only in the unphysical region with $c_s>1$.

\section{Introduction of the mean--field}

\subsection{Skyrme-like parametrization of the mean-field potential\label{Spar}}

The EOS obtained in preceding sections
takes  into account only short--range repulsive interactions of hadrons
and completely ignores the intermediate-range attractive interaction between baryons.
Due to this reason, this EOS does not describe the saturation property of
isospin--symmetric nuclear matter at low temperatures.
It~is well--known that such attractive interaction leads to the liquid--gas (LG) phase transition in
nuclear matter at tempera\-tures~$T\lesssim 10$\,MeV and chemical potentials
$\mu\sim\mu_0=m_N-E_B$\hsp, where \mbox{$E_B\simeq 16$ MeV} is the binding energy of
cold nuclear matter. This phase transition manifests itself
as the multifragmentation phenomenon in intermediate-energy nuclear reactions~\cite{Tra05}.

To account for the mean-field effects, we introduce an effective
potential \mbox{$U=U(n)$} which depends only on the baryon
density $n$ and does not depend on momenta of interacting baryons\hsp\footnote
{
For simplicity it is assumed that strange and nonstrange baryons have the same mean--field
potentials. On the other hand, in the considered case of chemically equilibrated matter with zero
strangeness, the medium-range interactions are not so important at high
tem\-pera\-tures~\mbox{$T\gtrsim 100$ MeV}, when the abundances of hyperons, antibaryons
and mesons become~\mbox{significant}.
}.
Then the baryon's single-particle energy can be obtained simply by adding~$U(n)$
to the kinetic energy. In this case
the partition function of the hadronic system can be calculated analytically~\cite{Ris91}.
Again, finite sizes of hadrons are taken into account by the volume reduction~(\ref{vrep})\hsp.
As the result, the following formulae for thermodynamic functions of the HP can be written
\begin{eqnarray}
&&\mu=\mu_K+U(n)\,,\label{fadd1}\\
&&P=P_K+P_f(n)\,,\label{fadd2}\\
&&\epsilon=\epsilon_K+\epsilon_f(n)\,.\label{fadd3}
\end{eqnarray}
The ''field'' contributions (marked by index $f$) to the energy density and pressure are found~as
\bel{fter}
\epsilon_f(n)=n\hsp U(n)-P_f(n)=\int\limits_0^{\,n} d\hspace*{.5pt}n_1U(n_1)\,.
\ee
The kinetic terms (marked by index $K$) in Eqs.~(\ref{fadd1})--(\ref{fadd3}) are functions
of $\mu_K,\mu_S,T$ calculated by using formulae of Sect.~\ref{HEOS} with the replacement
\mbox{$\mu\to\mu_K$}, \mbox{$P\to P_K$}, \mbox{$\epsilon\to\epsilon_K$}. The resulting expressions
for $P_K$ and $\epsilon_K$ are obtained from Eqs.~(\ref{pres})--(\ref{cceq}) and (\ref{ende}),
respectively. One can show that the total pressure $P=P(\mu,\mu_S,T)$ satisfies the
relation (\ref{ther}). The den\-si\-ties~$s, n$ and $n_S$ are given by
Eqs.~(\ref{entd})--(\ref{strd1}) where now
\bel{efcp1}
\widetilde{\mu_i}=B_i\hsp\mu_K+S_i\hsp\mu_S-v\hspace*{0.5pt} P_K\,.
\ee

\begin{figure*}[htb!]
\centerline{\includegraphics[width=0.7\textwidth]{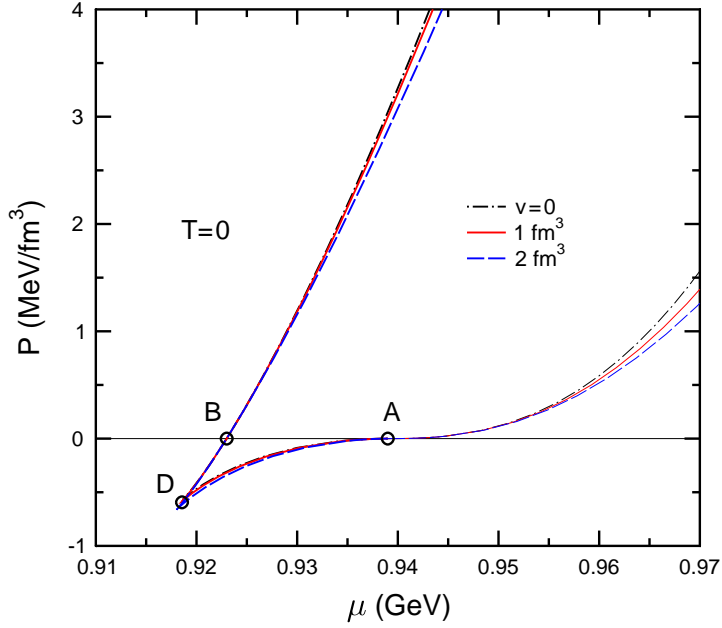}}
\caption{
Pressure of the baryonic system at $T=0$ as a function of $\mu$
for different values of the excluded volume $v$\hsp.
Thick and thin lines correspond to the cases \mbox{$U=U(n)$} and \mbox{$U=0$}, respectively.
The part $DA$ of the thick solid curve represents unstable states for $v=1$ fm$^3$.
}
\label{fig:m-p23}
\end{figure*}
In a spirit of the Skyrme approach~\cite{Sky56} we parametrize the mean--field potential in the form
\bel{mfpot}
U(n)=-\alpha\left(\frac{n}{n_0}\right)+\beta\left(\frac{n}{n_0}\right)^\gamma,
\ee
where $n_0$ is the saturation density of nuclear matter and
$\alpha,\beta,\gamma$ are density--independent parameters. In the following we fix $\gamma$
to a commonly used value $7/6$ and choose the remaining parameters from the
requirements $P=0$, \mbox{$\epsilon/n=\mu_0=923$} MeV at $n=n_0,\,T=0$\hsp. The
values of $\alpha,\beta$ as well as the incompressibility modulus $K=9\hsp\hsp\partial P/\partial n$
at the saturation point, calculated for different choices of $v$, are given in Table~\ref{tab4}.
\begin{table}[h!]
\caption{Parameters of the mean--field potential and the incompressibility modulus
of equilibrium nuclear matter for different values of the excluded volume\hsp.}
\label{tab4}
\vspace*{3mm}
\begin{tabular}{c|c|c|c}\hline
$v$\hsp, fm$^3$&0&1&2\\
\hline
$\alpha$\hsp, MeV&~352&~334&~297\\
$\beta$\hsp, MeV&~301&~277&~230\\
$K$\hsp, MeV&200&214&264\\
\hline
\end{tabular}
\end{table}

\vspace*{2mm}
\begin{figure*}[htb!]
\centerline{\includegraphics[width=0.7\textwidth]{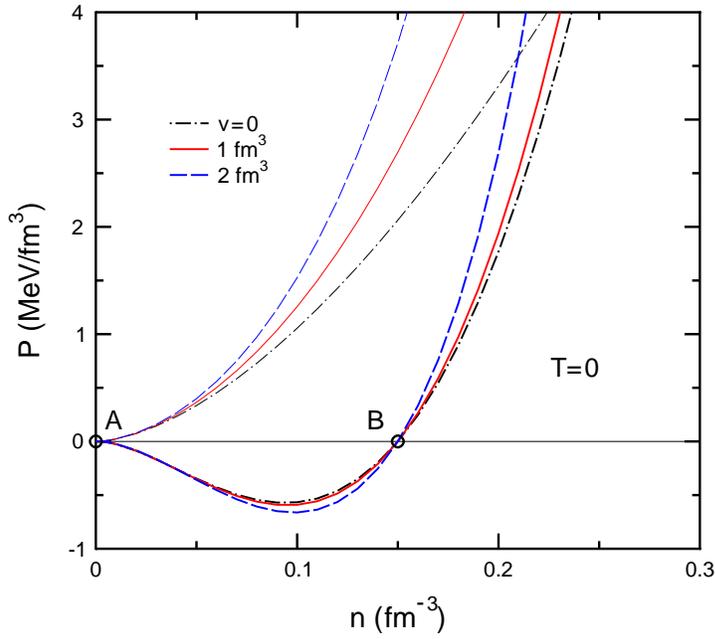}}
\vspace*{-4mm}
\caption{
Same as Fig.~\ref{fig:m-p23}, but for pressure as a function of the baryon density $n$\hsp.
}
\label{fig:n-p24}
\end{figure*}
\begin{figure*}[htb!]
\centerline{\includegraphics[width=0.7\textwidth]{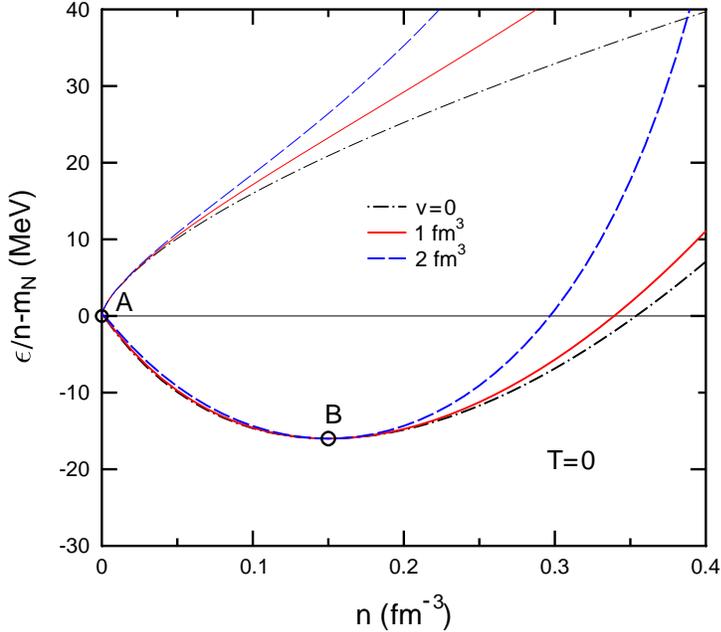}}
\vspace*{-4mm}
\caption{
Same as Fig.~\ref{fig:n-p24}, but for energy per baryon.
}
\label{fig:n-e25}
\end{figure*}
Figures \ref{fig:m-p23}--\ref{fig:n-e25} illustrate the properties of cold nuclear matter
as predicted by this model. Figure~\ref{fig:m-p23} shows pressure as a function of $\mu$ for
several values of the excluded volume $v$. Point $B$ marks the saturation point of cold nuclear matter.
The mean--field is responsible for the appearance of several branches
of pressure at \mbox{$\mu<\mu_A=m_N$}\hsp.
In the case $v=1$ fm$^3$ the branches $BA$ and~$DB$ describe the metastable states, while the branch $DA$
corresponds to the unstable states (see below). This behavior differs qualitatively from the case $U=0$
where only one branch is present at \mbox{$\mu\geqslant\mu_A$}\hsp.
Figures~\ref{fig:n-p24}--\ref{fig:n-e25} show pressure and energy per baryon
as functions of $n$\hsp. By construction, the saturation point has the same position for
all considered values of $v$\hsp.

\subsection{The liquid-gas phase transition}

\begin{figure*}[htb!]
\centerline{\includegraphics[width=0.7\textwidth]{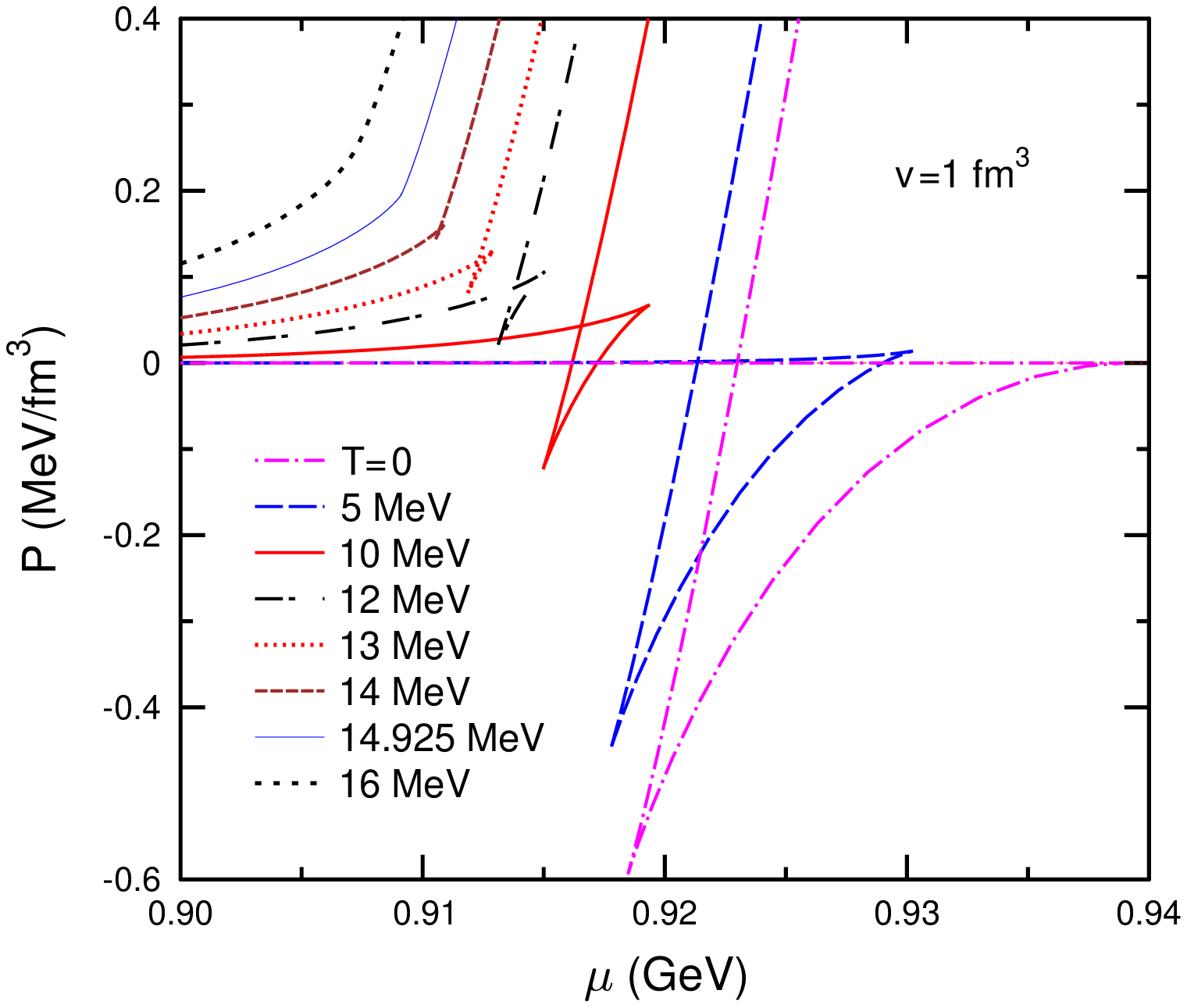}}
\caption{
Isotherms in the $\mu-P$ plane (\mbox{$v=1$ fm$^3$})\hsp.
}
\label{fig:m-p26}
\end{figure*}
\begin{figure*}[htb!]
\centerline{\includegraphics[width=0.7\textwidth]{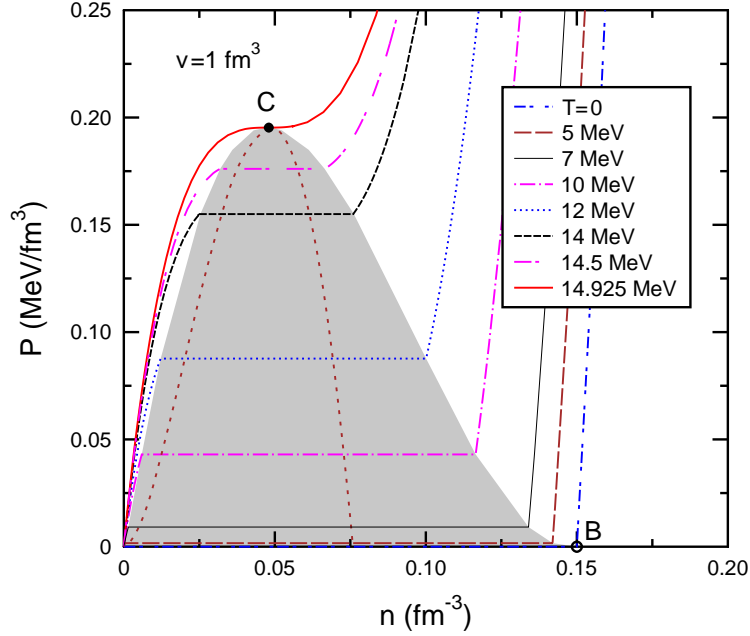}}
\caption{
Pressure isotherms as functions of baryon density (\mbox{$v=1$ fm$^3$})\hsp. The shaded area shows
the mixed phase region of the LG phase transition. The dotted line with maximum at
point $C$ is the boundary of the spinodal instability domain.
}
\label{fig:n-p27}
\end{figure*}
Now let us consider properties of hadronic matter at nonzero temperatures.
Figure~\ref{fig:m-p26} shows isotherms in the $\mu-P$ plane
calculated with $v=1$\,fm$^3$. In this case the existence of the critical point
is predicted at the temperature $T_c\simeq 14.925$\,MeV and the chemical
poten\-tial~\mbox{$\mu_c\simeq 909$\,MeV} (see Table~\ref{tab5}).
The isotherms with $T<T_c$ contain three branches of
pressure, similarly to the case $T=0$ discussed above. The phase transition points
correspond to intersections of metastable parts
of the isotherms. The branches with negative curvature correspond to unstable (spinodal) states.
Small density perturbations will grow exponentially in this region. This is a well-known
spinodal instability which leads to the separation of matter into dense and dilute
domains~\cite{Sau76}, characteristic of the LG phase transition. We would like to note here that
the~LG mixed phase can be represented by the ensemble of nuclear fragments of different sizes.
This ensemble can be well described by the statistical multifragmentation model~\cite{Bon95}
which also includes the EVC. The anomaly in the caloric curve, associated with the LG phase
transition, has been indeed observed in intermediate-energy heavy-ion collisions
by the ALADIN collaboration~\cite{Poc95}.

\begin{figure*}[htb!]
\centerline{\includegraphics[width=0.7\textwidth]{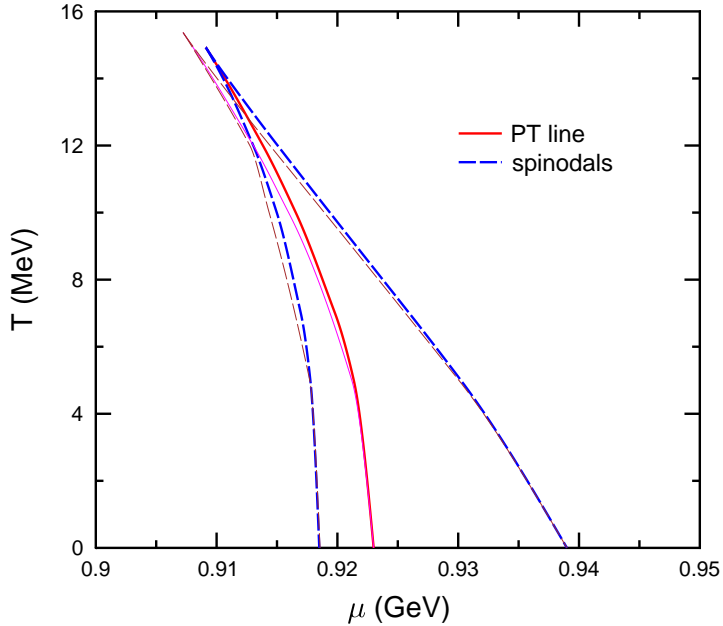}}
\caption{
Phase diagram of the LG phase transition in the $\mu-T$ plane (the solid line).
The dashed lines show boundaries of the spinodal region. Thin and thick lines correspond
to $v=0$ and $v=1\,\textrm{fm}^3$, respectively.
}
\label{fig:m-t28}
\end{figure*}

Figure~\ref{fig:n-p27} represents the isotherms in the $n-P$ plane.
The shaded region corresponds to the mixed phase, where the hadronic gas
(nucleons and light clusters) coexists with nuclear fragments (droplets of liquid).
Point $C$ is the critical point of the LG phase transition.
The parameters of the critical point for different values of $v$ are given in Table~\ref{tab5}.
\begin{table}[h!]
\caption{Characteristics of the critical point of the LG phase transition
at different values of the excluded volume\hsp.}
\label{tab5}
\vspace*{3mm}
\begin{tabular}{c|c|c|c}\hline
$v$\hsp, fm$^3$&0&1&2\\
\hline
$T_c$\hsp, MeV&~15.37&~14.93&~12.67\\
$\mu_c$\hsp, MeV&~907&~909&~915\\
$n_c$\hsp, fm$^{-3}$&0.048&0.048&0.039\\
$P_c$\hsp,  MeV/fm$^3$&0.198&0.195&0.139\\
\hline
\end{tabular}
\end{table}
One can see that their sensi\-ti\-vity to the excluded volume $v$ is rather weak.
This is also seen in Fig.~\ref{fig:m-t28} where we compare spinodals and the phase transition
lines for $v=0$ and~$v=1\,\textrm{fm}^3$. On the basis of these results we conclude
that characteristics of the LG phase transition in our model are similar to
predictions of other authors (see e.g.~\cite{Cho04}).

\begin{figure*}[htb!]
\centerline{\includegraphics[width=0.7\textwidth]{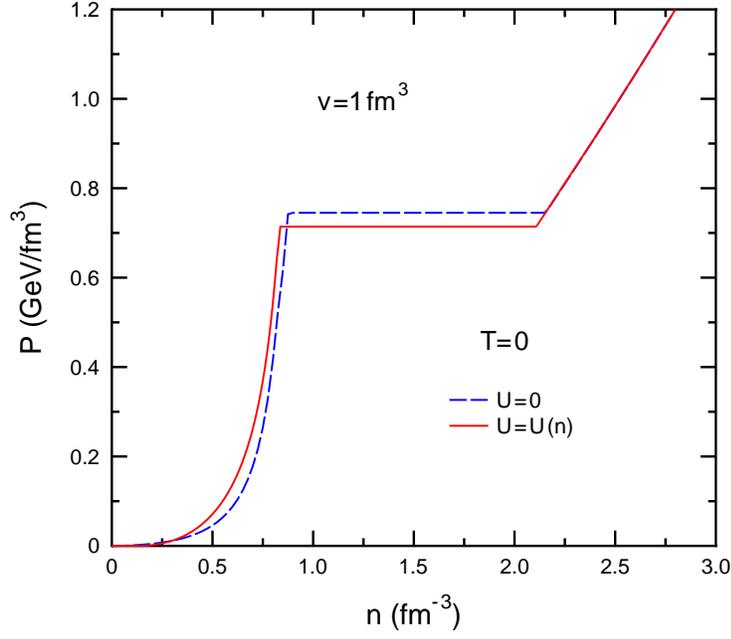}}
\caption{
Pressure as a function of baryon density at $T=0$\hsp.
The dashed and solid lines are calculated with and without the mean--field effects,
respectively ($v=1$\,fm$^{3}$)\hsp. Note change of the vertical scale compared
to Figs.~\ref{fig:n-p24}, \ref{fig:n-p27}.
}
\label{fig:n-p29}
\end{figure*}
\begin{figure*}[htb!]
\centerline{\includegraphics[width=0.7\textwidth]{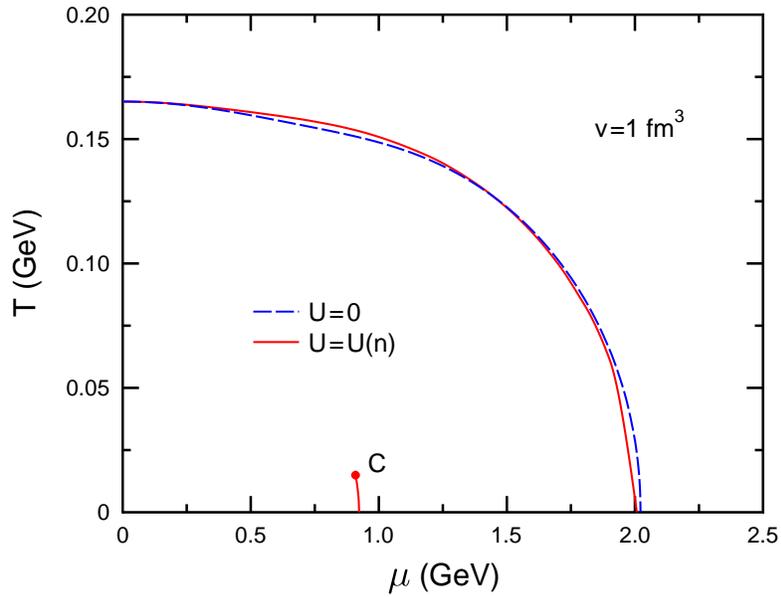}}
\vspace*{-4mm}
\caption{
Phase diagram of strongly interacting matter in the $\mu-T$ plane (\mbox{$v=1$\,fm$^{3}$})\hsp.
The dashed and solid lines are obtained from calculations with and without the mean--field effects,
respectively.
}
\label{fig:m-t30}
\end{figure*}
\begin{figure*}[htb!]
\centerline{\includegraphics[width=0.7\textwidth]{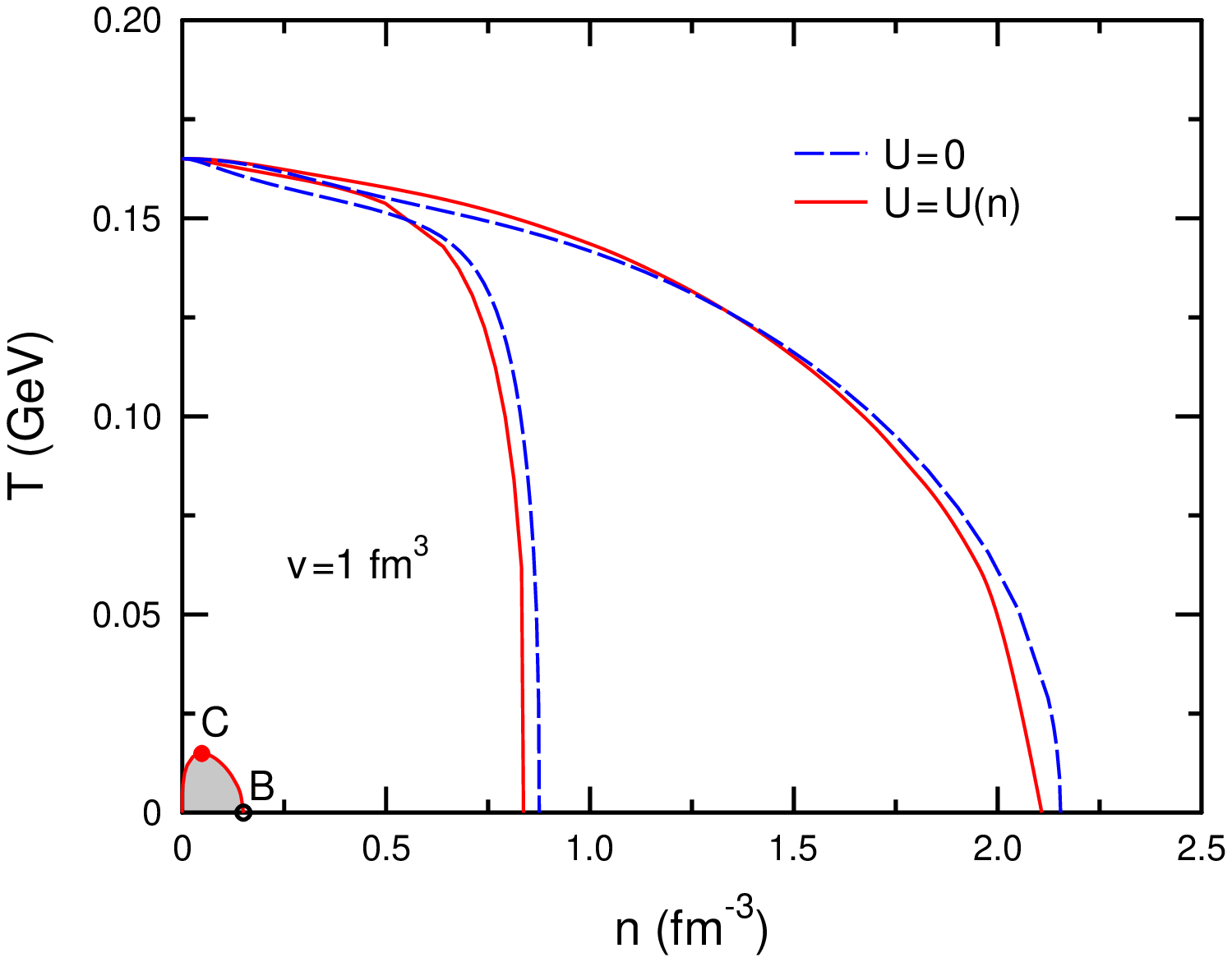}}
\vspace*{-4mm}
\caption{
Same as Fig.~\ref{fig:m-t30} but in the $n-T$ plane. Shading shows the region of the
LG phase transition.
}
\label{fig:n-t31}
\end{figure*}

\subsection{Phase diagram with two phase transitions}

Finally we present results for the full model where both the LG and deconfinement phase transitions
are  included. In the case $T=0$\hsp,
applying Eqs.~(\ref{fadd1})--(\ref{mfpot}) for a broad interval of baryon densities we get the results shown
in Fig.~\ref{fig:n-p29}. According to our calculations, in the region of the deconfinement phase transition
the corrections due to the mean--field interaction are rather small, of the order of 5\%\hsp.
Note that the mean--field potential leads to vanishing pressure for zero temperature states
with $n<n_0$ (see Fig.~\ref{fig:n-p27}).

Figures~\ref{fig:m-t30}--\ref{fig:n-t31},
represent the full phase diagram in the $\mu-T$ and $n-T$ planes for~$v=1$\,fm$^{3}$.
Compared to the calculation with $U=0$\hsp, a new phase transition line
starting at $\mu=\mu_0$ appears in Fig.~\ref{fig:m-t30}. As seen in Fig.~\ref{fig:n-t31},
the LG mixed phase (shown by shading) occupies a relatively small region of the
$n-T$ plane. According to Fig.~\ref{fig:n-t31} the borders of the quark--hadronic mixed phase are shifted
only slightly due to the mean--field effects.
The same conclusion is valid for the mixed phase boundaries and the $T=0$ line in the $n-\epsilon$
plane (cf.~Fig.~\ref{fig:n-e20}).

\section{Summary and discussion}

In this paper we have investigated the EOS of strongly interacting matter
in a phenomenological model taking into account the excluded volume effects in the hadronic phase.
The quark--gluon phase is described by the bag model with the lowest order perturbative
corrections in kinetic terms. Within this approach one can get only a first order phase transition.
The sensitivity of the phase diagram to the excluded volume $v$ has been investigated in details.
At small values of $v$ the phase diagram has unphysical behavior since the hadronic
phase becomes preferable at high temperatures. Only at $v\gtrsim 1$\,fm$^3$
the critical temperature of the deconfinement phase transition monotonously decreases
with baryon chemical potential $\mu$\hsp. Generally, the strength of the phase transition
increases with increasing $\mu$\hsp. Typical values of the baryon density in the mixed phase
are in the range $(5-10)\,n_0$ at $T\sim 100$\,MeV. Such a strong phase transition
should certainly lead to clear observable signals, e.g. the formation of quark--gluon droplets
at final stages of the relativistic nuclear collisions~\cite{Mis99,Tor08}.

The condition of zero net strangeness has been imposed in our calculations.
It leads to almost linear relation between the strange and baryon chemical potentials.
It is shown that the Bose condensation of $K^+$ mesons
is not possible at realistic values of $v$\hsp.
We have demonstrated the possibility of the strangeness--antistrangeness separation in the mixed phase.
This effect may help to produce clusters of strange matter like strangelets~\cite{Chi79,Far84}
or MEMOS~\cite{Sch92} in relativistic heavy-ion collisions.
We have calculated the adiabatic trajectories ($S/B=\textrm{const}$) and found
that they have a zigzag-like shape in the mixed phase region.
According to our analysis, sometimes the model predicts acausal states (\mbox{$c_s>1$}), but they
lie outside the region reachable in ultrarelativistic heavy--ion
collisions. By implementing the mean--field potential
for baryons we have described simultaneously the liquid-gas and deconfinement
phase transitions. However, properties of the quark-hadron mixed phase are only slightly
influenced by the mean--field effects.

In the future we are going to generalize this model by
introducing different excluded volumes for different hadronic species. Some attempts
in this direction have been already made in Refs.~\cite{Yen97,Gor05,Gor99}.
Our present approach does not take into account modifications of particle
properties in a dense medium, although hadronic masses and radii
may significantly change as compared to their vacuum values. This problem should be also studied
in the future.

\begin{acknowledgments}
The authors thank M.I.~Gorenstein for numerous fruitful discussions. We are also
grateful to M. Bleicher, M. Hauer, Yu.B.~Ivanov, A.V.~Merdeev,
V.N.~Russkikh and G.~Torrieri for the interest to this work.
L.M.S. acknowledges the kind hospitality and financial support from FIAS.
This work was supported in part by the GSI, the DFG grant~436
RUS~\mbox{113/957/0--1} (Germany), the grants NS--3004.2008.2 and
RFFI 09-02-91331 (Russia).
\end{acknowledgments}

\appendix
\setcounter{equation}{0}
\renewcommand{\theequation}{A.\arabic{equation}}
\section{Thermodynamic functions of hadronic system}

In the case of fermions ($B_i=\pm 1$) one can represent the integrals
in~\re{therf} in the form~\cite{Lan80} which makes easier their numerical calculation
at low temperatures:
\bel{therf1}
\hspace*{-2pt}\left(\begin{array}{c}
\hspace*{-2pt}\widetilde{\epsilon}_i\hspace*{-2pt}\\P_i\\
\widetilde{n}_i
\end{array}\right)\hspace*{-3pt}=\frac{g_i}{2\hsp\pi^2}
\int\limits_{m_i}^{\infty}d\epsilon\hspace*{1.5pt}
\frac{\sqrt{\epsilon^2-m_i^2}}{\ds e^{\frac{\ds|\epsilon-\widetilde{\mu}_i|}
{\raisebox{-5pt}{$T$}}}+1}\hspace*{1.5pt}
\textrm{sgn}\hspace*{.5pt}(\epsilon-\widetilde{\mu}_i)
\hspace*{-1pt}\left(\begin{array}{c}
\epsilon^2\\
\hspace*{-4pt}\frac{1}{3}(\epsilon^2-m_i^2)\hspace*{-4pt}\\
\epsilon
\end{array}\right)\hspace*{-2pt}+\hsp
\frac{g_i}{2\hsp\pi^2}\hspace*{-1pt}\left(\begin{array}{c}
\frac{p_F^{\hsp 4}}{4}\hsp\psi\hspace*{-2pt}\left(\frac{m_i}{p_F}\right)\\
\hspace*{-4pt}\frac{\widetilde{\mu}_i p_F^{\hsp 3}}{3}-\frac{p_F^{\hsp 4}}
{4}\hsp\psi\hspace*{-2pt}\left(\frac{m_i}
{p_F}\right)\hspace*{-5pt}\\
\frac{p_F^{\hsp 3}}{3}
\end{array}\right)\hspace*{-2pt}\Theta\hsp(\widetilde{\mu}_i-m_i)\hsp.
\ee
Here $p_F=\sqrt{\widetilde{\mu}_i^{\hsp 2}-m_i^2}\hsp,\,
\Theta\hsp (x)=\frac{1}{2}\hsp(1+\textrm{sgn}\hsp x)$ and $\psi\hsp (x)$ is defined as
\bel{psif}
\psi(x)=4\int\limits_0^1 d\hspace*{.5pt}t\hsp t^{\hsp 2}\sqrt{t^2+x^2}=
\left(1+\frac{x^2}{2}\right)\hspace*{-1pt}\sqrt{1+x^2}-
\frac{x^4}{2}\ln{\frac{1+\sqrt{1+x^2}}{x}}\,.
\ee
The integrals in the first term of Eq.~(\ref{therf1}) vanish at $T\to 0$\hsp.
They were calculated numerically using the Newton-Cotes method.

In the case of mesons ($B_i=0$) we calculate integrals in~\re{therf} by representing them as
series of modified Bessel functions:
\bel{therb}
\left(\begin{array}{c}
\widetilde{\epsilon}_i\\P_i\\
\widetilde{n}_i
\end{array}\right)=\frac{g_i\hsp m_i^3}{2\hsp\pi^2}\sum\limits_{l=1}^{\infty}
\exp{\left(\frac{\ds\widetilde{\mu}_i\hsp l}
{\ds T}\right)}\hspace*{-1pt}\left(\begin{array}{c}
\frac{\ds m_i}{\ds x}\hspace*{-2pt}\left(K_1+\frac{\ds 3K_2}{\ds x}\right)\\
\raisebox{1pt}{$\frac{\ds m_iK_2}{\ds\raisebox{-4pt}{$x^2$}}$}\\
\raisebox{-2pt}{$\frac{\ds K_2}{\ds x}$}
\end{array}\right),
\ee
where $x=m_i l/T$ and $K_n=K_n(x)$ is the MacDonald function of the $n$--th order.
These series converge if $\widetilde{\mu}_i<m_i$ (see first footnote on page~\pageref{fref1}).



\vspace*{-3mm}
\begin{thebibliography}{00}

\bibitem{Kar01}
        F. Karsch, E. Laermann, and A. Peikert,
        Nucl. Phys. B \textbf{605}, 579 (2001).

\bibitem{Gyu04}
        M. Gyulassy and L. McLerran,
        Nucl. Phys. A \textbf{750}, 30 (2004).

\bibitem{Phi07}
       P. de Forcrand and O. Philipsen, JHEP \textbf{01}, 077 (2007).

\bibitem{Alf98}
       M.G. Alford, K. Rajagopal, and F. Wilczek,
       Phys. Lett. B \textbf{422}, 247 (1998).

\bibitem{Sca01}
       O. Scavenius, A. Mocsy, I.N. Mishustin, and D.H. Rischke,
       Phys. Rev. C \textbf{64}, 045202 (2001).

\bibitem{Ars07}
       I.C. Arsene, L.V. Bravina, W. Cassing, Yu.B. Ivanov, A. Larionov,
       J. Randrup, V.N. Russkikh, V.D. Toneev, G. Zeeb, and D. Zschiesche,
       Phys. Rev. C \textbf{75}, 034902 (2007).

\bibitem{Afa02}
        S.V. Afanasiev et al. (NA49 Collaboration),
        Phys. Rev. C \textbf{66}, 054902 (2002).

\bibitem{Alt08}
        C. Alt et al. (NA49 Collaboration),
        Phys. Rev. C \textbf{77}, 024903 (2008).

\bibitem{Mis02}
        I.N. Mishustin, L.M. Satarov, H. St\"ocker, and W. Greiner,
        Phys. Rev. C \textbf{66}, 015202 (2002).

\bibitem{Hei86}
        U. Heinz, P.R. Subramanian, H. St\"ocker, and W. Greiner,
        J. Phys. G \textbf{12}, 1237 (1986).

\bibitem{Wal74}
        J.D. Walecka, Ann. Phys. \textbf{83}, 491 (1974).

\bibitem{Lan80} L.D. Landau and E.M. Lifshitz, \textit{Statistical
                physics}, Pergamon Press, 1980.

\bibitem{Hag80}
       R. Hageddorn and J. Rafelski, Phys. Lett. \textbf{97B}, 136 (1980).

\bibitem{Kar80}
       F. Karsch and H. Satz,
       Phys. Rev. D \textbf{21}, 1168 (1980).

\bibitem{Kap81}
       J.I. Kapusta,
       Phys. Rev. D \textbf{23}, 2444 (1981).

\bibitem{Ris91}
        D.H. Rischke, M.I. Gorenstein, H. St\"ocker, and W. Greiner,
        Z. Phys. C \textbf{51}, 485 (1991).

\bibitem{Ven92}
       R. Veugopalan and M. Prakash,
       Nucl. Phys. A \textbf{546}, 718 (1992).

\bibitem{Yen97}
        G.D. Yen, M.I. Gorenstein, W. Greiner, and S.N. Yang,
        Phys. Rev. C \textbf{56}, 2210 (1997).

\bibitem{Whe04}
        S. Wheaton and J. Cleymans, hep-ph/0407174.

\bibitem{Gre87}
        C. Greiner, P. Koch, and H. St\"ocker, Phys. Rev. Lett.
        \textbf{58}, 1825 (1987).

\bibitem{Gor05}
        M.I. Gorenstein, M. Ga\'zdzicki, and W. Greiner,
        Phys. Rev. C \textbf{72}, 024909 (2005).


\bibitem{PDG08}
        Particle Data Group, C. Amsler et al.,
        Phys. Lett. B \textbf{667}, 1 (2008).

\bibitem{Kap86}
        D.B. Kaplan and A.E. Nelson, Phys. Lett. B \textbf{175}, 57 (1986).

\bibitem{Sch94}
        \mbox{J. Schaffner, A. Gal, I.N. Mishustin, H. St\"ocker, and
        W. Greiner,}\\ Phys. Lett. B \textbf{334}, 268 (1994).

\bibitem{Sch97}
         J. Schaffner, J. Bondorf, and I.N. Mishustin,
         Nucl. Phys. A \textbf{625}, 325 (1997).

\bibitem{Zak05}
        I. Zakout, W. Greiner, and H.R. Jaqaman,
        Nucl. Phys. A \textbf{759}, 201 (2005).

\bibitem{Cle93}
        J. Cleymans and H. Satz, Z. Phys. C \textbf{57}, 135 (1993).

\bibitem{Bra96}
        P. Braun--Munzinger, and J. Stachel,  Nucl. Phys. A \textbf{606}, 320 (1996).

\bibitem{And08}
        A. Andronic, P. Braun--Munzinger, and J. Stachel,
        ArXiv: 0812.1186 [nucl-th].

\bibitem{Ahl00}
        L. Ahle et al. (E866/E917 Collaboration),
        Phys. Lett. B \textbf{476}, 1 (2000).

\bibitem{Abe08}
        B.I. Abelev et al. (STAR Collaboration),
        ArXiv: 0808.2041 [nucl-ex].

\bibitem{Pink02}
        C. Pinkenburg et al. (E895 Collaboration), Nucl. Phys. A \textbf{698}, 495 (2002).

\bibitem{Alb02}
        S. Albergo et al. (E896 Collaboration), Phys. Rev. Lett. \textbf{88}, 062301 (2002).

\bibitem{Bec04}
        F. Becattini, M. Ga\'zdzicki, A. Ker\"anen, J. Manninen, and R. Stock,
        Phys. Rev. C \textbf{69}, 024905~(2004).

\bibitem{Iva05}
        Yu.B.~Ivanov, A.S.~Khvorostukhin, E.E.~Kolomeitsev, V.V.~Skokov, V.D.~Toneev,\\
        and D.N.~Voskresensky, Phys. Rev. C \textbf{72}, 025804 (2005).

\bibitem{Sat07}
        L.M. Satarov, I.N. Mishustin, A.V. Merdeev, and H. St\"ocker,
        Phys. Rev. C \textbf{75}, 024903 (2007).

\bibitem{Sat07b}
        \mbox{L.M. Satarov, I.N. Mishustin, A.V. Merdeev, and H. St\"ocker,
        Yad. Fiz. \textbf{70}, 1822 (2007)}
        [Phys. Atom. Nucl. \textbf{70}, 1773 (2007)\hsp]\hsp.

\bibitem{Tea01}
        D. Teaney, J. Lauret, and E.V. Shuryak,
        Phys. Rev. Lett. \textbf{86}, 4783 (2001).

\bibitem{Lan87}
         L.D. Landau and E.M. Lifshitz, \textit{Fluid
         Mechanics}, Pergamon Press, 1987.

\bibitem{And06}
        A. Andronic, P. Braun--Munzinger, and J. Stachel,
        Nucl. Phys. A \textbf{772}, 167 (2006).


\bibitem{Mis01}
         I.N. Mishustin, L.M. Satarov, H. St\"ocker, and W. Greiner,
         Yad. Fiz. \textbf{64}, 866 (2001)\\
         \mbox{[Phys. Atom. Nucl. \textbf{64}, 802 (2001)\hsp]}.

\bibitem{Cle98}
        J. Cleymans and K. Redlich, Phys. Rev. Lett. \textbf{81}, 5284 (1998).

\bibitem{Pae01} K. Paech, M. Reiter, A. Dumitru, H. St\"ocker, and
                W. Greiner, Nucl. Phys. A \textbf{681}, 41 (2001).

\bibitem{Tra05}
        W. Trautmann, Nucl. Phys. A \textbf{752}, 407 (2005).

\bibitem{Sky56}
        T.H.R. Skyrme, Phil. Mag. \textbf{1}, 1043 (1956);
        Nucl. Phys. \textbf{9}, 615 (1959).

\bibitem{Sau76}
        G. Sauer, H. Chandra, and U. Mosel, Nucl. Phys. A \textbf{264}, 221 (1976).

\bibitem{Bon95}
        J.P. Bondorf, A.S. Botvina, A.S. Iljinov, I.N. Mishustin, and K. Sneppen,
        Phys. Rep. \textbf{257}, 133 (1995).

\bibitem{Poc95}
        J. Pochodzalla et al., Phys. Rev. Lett. \textbf{75}, 1040 (1995).


\bibitem{Cho04}
        P. Chomaz, M. Colonna, and J. Randrup, Phys. Rep. \textbf{389}, 263 (2004).

\bibitem{Mis99}
        I.N. Mishustin, Phys. Rev. Lett. \textbf{82}, 4779 (1999).

\bibitem{Tor08}
        G. Torrieri, B. Tomasik, and I.N. Mishustin,
        Phys. Rev. C \textbf{77}, 034903 (2008).

\bibitem{Chi79}
        S.A. Chin and A.K. Kerman, Phys. Rev. Lett. \textbf{43}, 1292 (1979).

\bibitem{Far84}
        E. Farhi and R.L. Yaffe, Phys. Rev. D \textbf{30}, 2379 (1984).

\bibitem{Sch92}
       J. Schaffner, C. Greiner, and H. St\"ocker, Phys. Rev. C \textbf{46}, 322 (1992).

\bibitem{Gor99}
        M.I. Gorenstein, A.P. Kostyuk, and Ya.D. Krivenko,
        J. Phys. G \textbf{25}, L75 (1999).

\end{thebibliography}
\end{document}